\documentclass[aps,reprint,twocolumn,superscriptaddress,citeautoscript,amsmath,amssymb,floatfix,longbibliography,prl,]{revtex4-2}
\usepackage{graphicx}
\usepackage{times}
\usepackage{bm}
\usepackage{color}
\usepackage{gensymb}
\usepackage{dcolumn}
\usepackage{xfrac}

\newcolumntype{d}[1]{D{.}{.}{#1}}

\begin{document}
	
\title{Carrier Tuning of Stoner Ferromagnetism in ThCr$_{\bm{2}}$Si$_{\bm{2}}$-Structure Cobalt Arsenides}

\author{B.~G.~Ueland}
\email{bgueland@ameslab.gov}
\affiliation{Ames Laboratory, U.S. DOE, Iowa State University, Ames, Iowa 50011, USA}
\affiliation{Department of Physics and Astronomy, Iowa State University, Ames, Iowa 50011, USA}

\author{Santanu~Pakhira}
\affiliation{Ames Laboratory, U.S. DOE, Iowa State University, Ames, Iowa 50011, USA}
\affiliation{Department of Physics and Astronomy, Iowa State University, Ames, Iowa 50011, USA}

\author{Bing~Li}
\affiliation{Ames Laboratory, U.S. DOE, Iowa State University, Ames, Iowa 50011, USA}
\affiliation{Department of Physics and Astronomy, Iowa State University, Ames, Iowa 50011, USA}

\author{A.~Sapkota}
\altaffiliation[Present Address:  ]{Condensed Matter Physics and Materials Science Division, Brookhaven National Laboratory, Upton, NY 11973, USA}
\affiliation{Ames Laboratory, U.S. DOE, Iowa State University, Ames, Iowa 50011, USA}
\affiliation{Department of Physics and Astronomy, Iowa State University, Ames, Iowa 50011, USA}

\author{N.~S.~Sangeetha}
\altaffiliation[Present Address:  ]{Institute for Quantum Materials and Technologies, Karlsruhe Institute of Technology, 76021 Karlsruhe, Germany}
\affiliation{Ames Laboratory, U.S. DOE, Iowa State University, Ames, Iowa 50011, USA}

\author{T.~G.~Perring}
\affiliation{ISIS Neutron and Muon Source, STFC Rutherford Appleton Laboratory, Didcot, Oxon OX11 0QX, UK}

\author{Y.~Lee}
\affiliation{Ames Laboratory, U.S. DOE, Iowa State University, Ames, Iowa 50011, USA}

\author{Liqin Ke}
\affiliation{Ames Laboratory, U.S. DOE, Iowa State University, Ames, Iowa 50011, USA}

\author{D.~C.~Johnston}
\affiliation{Ames Laboratory, U.S. DOE, Iowa State University, Ames, Iowa 50011, USA}
\affiliation{Department of Physics and Astronomy, Iowa State University, Ames, Iowa 50011, USA}

\author{R.~J.~McQueeney}
\email{rmcqueen@ameslab.gov}
\affiliation{Ames Laboratory, U.S. DOE, Iowa State University, Ames, Iowa 50011, USA}
\affiliation{Department of Physics and Astronomy, Iowa State University, Ames, Iowa 50011, USA}

\date{\today}

\begin{abstract}
	CaCo$_{2-y}$As$_2$ is an unusual itinerant magnet with signatures of extreme magnetic frustration.  The conditions for establishing magnetic order in such itinerant frustrated magnets, either by reducing frustration or increasing electronic correlations, is an open question.  Here we use results from inelastic neutron scattering and magnetic susceptibility measurements and density functional theory calculations to show that hole doping in Ca(Co$_{1-x}$Fe$_{x}$)$_{2-y}$As$_{2}$ suppresses magnetic order by quenching the magnetic moment while maintaining the same level of magnetic frustration. The suppression is due to tuning the Fermi energy away from a peak in the electronic density of states originating from a flat conduction band.  This results in the complete elimination of the magnetic moment by $x\approx0.25$, providing a clear example of a Stoner-type transition.
\end{abstract}

\maketitle


Iron and cobalt pnictide metals harbor weak to moderate magnetism driven by features in their electronic-band structure lying close to the Fermi energy $E_{\text{F}}$.  Tuning the chemical composition of such materials has resulted in intriguing properties related to the underlying magnetism including non-Fermi-liquid behavior \cite{Sangeetha_2019}, magnetic glassiness \cite{Pakhira_2020}, electronic nematicity \cite{Fernandes_2012}, and unconventional superconductivity \cite{Johnston_2010,Canfield_2010, Stewart_2011}.  While often discussed using a local-moment description \cite{Diallo_2010,Sapkota_2017,Jayasekara_2013}, it is clear that the itinerant nature of the magnetism in these compounds is essential for facilitating the tunability of these phenomena.  More generally, compared to our knowledge of local-moment magnetism, our understanding of itinerant magnetism is limited by the relatively poorer experimental representation of purely itinerant-moment systems \cite{Santiago_2017}.  In this work, we report the direct observation of quenching of the magnetic moment in a Co pnictide by a Stoner-type transition \cite{Stoner_1938}.

\begin{figure}[]
	\centering
	\includegraphics[width=1.0\linewidth]{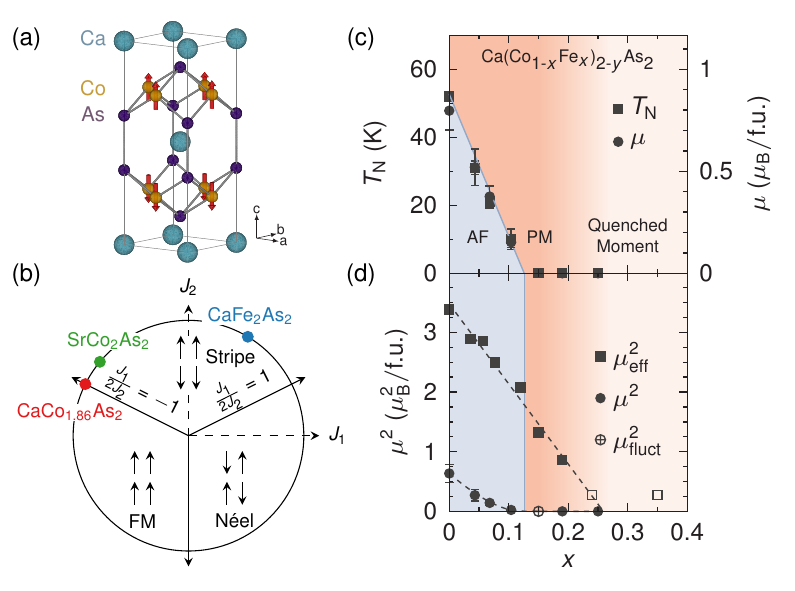}
	\caption{  \label{Fig:structure_J1J2} (a) The unit cell (space group $I\frac{4}{m}mm$) of CaCo$_{1.86}$As$_{2}$ with its A-type antiferromagnetic (AF) structure shown by red arrows; $a=b=3.9906(1)$~\AA\ and $c=10.280(1)$~\AA\ at $T=300$~K \cite{Quirinale_2013,Momma_2008}. (b) Phase diagram for the $J_1$-$J_2$  Heisenberg model on a square lattice. FM corresponds to the A-type order of CaCo$_{1.86}$As$_{2}$. (c) Magnetic phase diagram for Ca(Co$_{1-x}$Fe$_{x}$)$_{2-y}$As$_{2}$ \cite{Jayasekara_2017} showing the N\'{e}el temperature $T_{\text{N}}$ and ordered magnetic moment $\mu$ versus $x$.   PM is paramagnetic. The ``quenched moment'' region has neither static nor dynamic spin correlations.   (d) Plots of the $x$-dependence of $\mu^2$, the square of the fluctuating moment $\mu_{\text{fluct}}^2$, and the square of the effective  moment $\mu_{\text{eff}}^2$.  $\mu_{\text{eff}}^2$ for $x=0$ is from Ref.~[\onlinecite{Anand_2014}].   Open squares indicate values for which the modified Curie-Weiss fits used to determine $\mu_{\text{eff}}^2$ are not valid \cite{SM}.  Lines are guides to the eye.}
\end{figure}

The ThCr$_2$Si$_2$-type ($122$) pnictide  CaCo$_{2-y}$As$_{2}$, where $y$ corresponds to vacancies on the Co site has the crystal structure shown in Fig.~\ref{Fig:structure_J1J2}(a) \cite{Quirinale_2013, Momma_2008} which is closely matched to the Fe-pnictide superconductors \cite{Diallo_2009,Diallo_2010,Harnagea_2011,Hu_2012}. Contemporary studies of CaCo$_{2-y}$As$_{2}$ were initially aimed at discovering the conditions necessary to create a superconducting state similar to that found in the Fe-based pnictides.  However, its A-type antiferromagnetic (AF) order [shown in Fig.~\ref{Fig:structure_J1J2}(a)] was found to be quite intriguing, exhibiting  ferromagnetic (FM) Co layers with evidence for extreme magnetic frustration \cite{Sapkota_2017,SM} and signatures of itinerant magnetism \cite{Mao_2018,Quirinale_2013, Anand_2014,Jayasekara_2017}.

Extreme frustration was found in  CaCo$_{1.86}$As$_{2}$ via inelastic neutron scattering (INS) measurements made below the N\'{e}el temperature of $T_{\text{N}}=52(1)$~K \cite{Quirinale_2013, Anand_2014, Jayasekara_2017}.  These data show quasi-one-dimensional ($1$D) spin fluctuations dominated by the FM Co layers instead of well-defined spin waves \cite{Sapkota_2017}.  As explained below, describing this behavior using a local-moment (Heisenberg) model places the compound at the border between FM and stripe-AF ordering which indicates extreme frustration. On the other hand, CaCo$_{1.86}$As$_{2}$ exhibits a weak ordered magnetic moment  of $\mu=0.80(9)~\mu_{\text{B}}/\text{f.u.}$ \cite{Jayasekara_2017}, temperature-independent contributions to the magnetic susceptibility $\chi$, and a somewhat large Sommerfeld coefficient of $\gamma=27(1)~\text{mJ}/\text{mol-K}^2$ \cite{Anand_2014} which all point to its magnetism being itinerant.

In general, frustrated and itinerant magnetic systems each have different conditions for establishing magnetic order. In the former case, some relief from frustration, for example, by modifications of the exchange constants within a Heisenberg model, is needed. The latter case can occur by exceeding the Stoner criterion $\alpha_0 = \rho(E_{\text{F}})I>1$, where $\rho(E_{\text{F}})$ is the density of electronic states at the Fermi energy and $I$ is the effective Coulomb repulsion \cite{Stoner_1938,Moriya_1985,Takahashi_2013}.  Here we address which phenomenon is operable in CaCo$_{2-y}$As$_{2}$ and present a compelling example of Stoner ferromagnetism in the presence of extreme frustration.  Our INS and $\chi$ data reveal the complete elimination of the fluctuating magnetic moment $\mu_{\text{fluct}}$ at $x=0.25$ without any measurable change to the magnetic frustration.  Our density functional theory (DFT) calculations establish that Fe substitution dopes holes into the system and shifts $E_{\text{F}}$ away from a flat electronic band that creates a large peak in $\rho(E)$. \cite{Tasaki_1998,Mao_2018}.  We conclude that a Stoner-type transition is induced by hole doping.

Without getting into the microscopic details of the exchange pathways, which is a subject of some debate, we note that the the $J_1$-$J_2$ Heisenberg model for a square lattice, with exchange constants $J_1$ and $J_2$ between nearest-neighbor (NN) and next-nearest-neighbor spins, respectively, has been used for many $122$-pnictides \cite{Dai_2015, SM}.  This includes CaCo$_{1.86}$As$_{2}$ \cite{Sapkota_2017} where the effective exchange-interaction strength between transition metal layers is much weaker than the effective interactions within the planes  \cite{Sapkota_2017}.

Within this model, the quasi-$1$D spin fluctuations in  CaCo$_{1.86}$As$_{2}$ give the ratio  $\eta=J_1/(2J_2)=-1.03(2)\approx-1$ \cite{Sapkota_2017}.  This indicates extreme frustration because it locates the compound at the border between the FM and stripe-AF phases in Fig.~\ref{Fig:structure_J1J2}(b).  CaFe$_{2}$As$_{2}$, on the other hand, lies in the stripe region with an AF $J_1$ and exhibits stripe-AF order \cite{Diallo_2009,Diallo_2010}, whereas the stripe-AF spin fluctuations in paramagnetic (PM) SrCo$_2$As$_2$ require a smaller value of $\eta$.  This suggests that the exchange constants and, hence, magnetic frustration in these cobalt arsenides is tunable \cite{Jayasekara_2013,Li_2019}.  Such tunability, which in principle might be possible by carrier doping, offers the enticing prospect of finding a quantum phase transition \cite{Brando_2016} and spin-liquid ground states \cite{Savary_2016,Zhou_2017}.  More discussion of the Stoner and  $J_1$-$J_2$ models is given in the Supplemental Material (SM) \cite{SM}.

In this respect, it is interesting to study the evolution of the spin fluctuation spectrum of Ca(Co$_{1-x}$Fe$_{x}$)$_{2-y}$As$_{2}$ since Fe substitution (nominal hole doping) suppresses magnetic order by sending both $T_{\text{N}}$ and $\mu\rightarrow0$ at $x=0.12(1)$ \cite{Jayasekara_2017}.  Further, a large body of work on the $A$(Fe$_{1-x}$Co$_{x}$)$_{2}$As$_{2}$, $A=$~Ca, Sr, or Ba, high-$T_{\text{c}}$ superconductors and related compounds shows that the ratio of Co to Fe rigidly shifts $E_{\text{F}}$ albeit with some small level of band broadening due to disorder \cite{Canfield_2010, Johnston_2010,  Dai_2012, Sangeetha_2019, LiY_2019b}.  Thus, a careful study of Ca(Co$_{1-x}$Fe$_{x}$)$_{2-y}$As$_{2}$ can address fundamental questions regarding  the origin of its collective magnetism  and whether critical compositions lead to strong quantum fluctuations and novel properties. 

Plate-like single crystals of Ca(Co$_{1-x}$Fe$_{x}$)$_{2-y}$As$_{2}$ were solution grown using Sn flux and their compositions were measured using  energy-dispersive x-ray spectroscopy.  INS data were collected at $T=5.5$~K for a $2.1$~g coaligned single-crystal sample of Ca(Co$_{0.85}$Fe$_{0.15}$)$_{2}$As$_{2}$ using the MERLIN spectrometer at the ISIS Neutron and Muon Source at the Rutherford Appleton Laboratory \cite{MERLIN_2018}.  Measurements were made with $\mathbf{c}$ fixed parallel to the incident neutron beam which links the $L$ reciprocal lattice direction to $E$.  $\chi(T)$ was determined using using a Quantum Design, Inc., SQUID magnetometer. The powder average of $\chi$ [$\chi_{\text{ave}}=(2/3)\chi_{ab}+(1/3)\chi_{c}$] was  found by measuring $\chi$ perpendicular ($\chi_{ab}$) and parallel ($\chi_{c}$) to $\mathbf{c}$.  DFT calculations were performed using the full-potential linear-augmented-plane-wave (FP-LAPW) method \cite{wien2K} with the generalized gradient approximation (GGA) \cite{Perdew_1996}.  Further details are given in the SM \cite{SM}.   Potential effects of chemical disorder on the magnetic order are discussed in Ref.~[\onlinecite{Jayasekara_2017}].  Since Ca(Co$_{1-x}$Fe$_{x}$)$_{2-y}$As$_{2}$ exists in the collapsed-tetragonal phase for $x\lesssim0.5$ \cite{Jayasekara_2017,Hoffmann_1985,Reehuis_1990,Reehuis_1998}, and CaFe$_2$As$_2$ is nonmagnetic in the collapsed-tetragonal phase \cite{Kreyssig_2008, Goldman_2009, Soh_2013}, we expect Fe to be nonmagnetic for the values of $x$ studied here.


\begin{figure}[]
	\centering
	\includegraphics[width=1.0\linewidth]{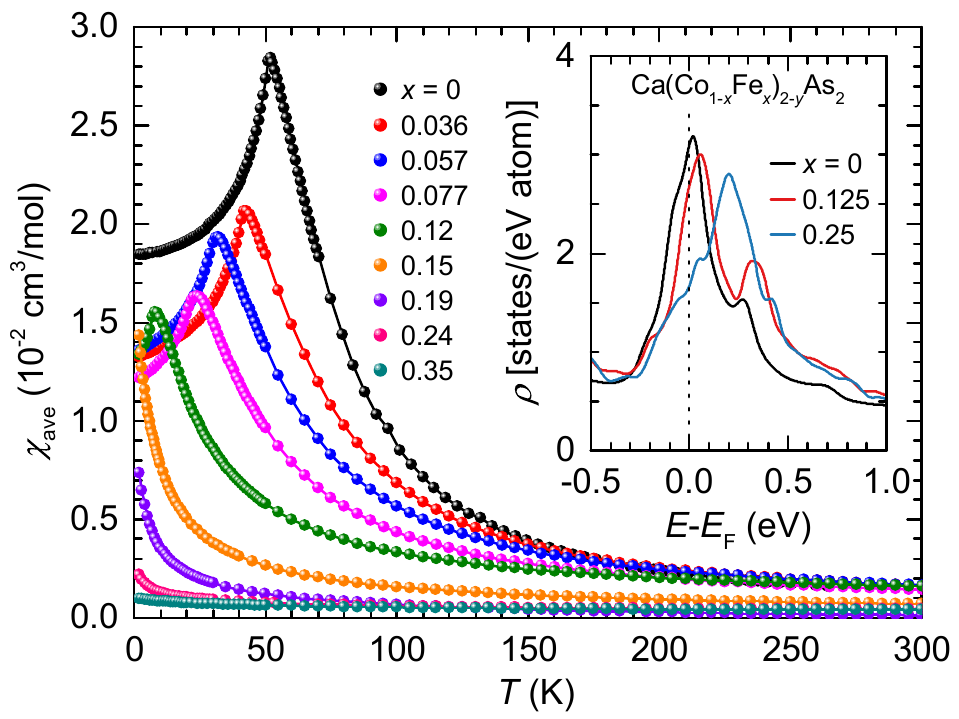}
	\caption{  \label{Fig:chiave_allx} The powder average of the static magnetic susceptibility $\chi_{\text{ave}}$ versus $T$ for various $x$.  The inset shows the partial electronic density of states per transition metal atom $\rho$ for $x=0$, $0.125$ and $0.25$ and $y=0$.  $E_{\text{F}}$ is the Fermi energy.}
\end{figure}

We begin the presentation of our results by showing the overall suppression of $\chi(T)$ with increasing $x$  in Fig.~\ref{Fig:chiave_allx}. Peaks occur in $\chi$ near $T_{\text{N}}$ for samples exhibiting A-type AF order.  We quantify the  suppression of $\chi$ with $x$ by determining the effective magnetic moment $\mu_{\text{eff}}(x)$ per formula unit through fitting a modified Curie-Weiss law to $\chi_{\text{ave}}^{-1}(T)$  as shown in the SM \cite{SM,Kittel_2004,Ashcroft_2020}.  Figure~\ref{Fig:structure_J1J2}(d) shows that $\mu_{\text{eff}}^2$ decreases with increasing $x$, remaining finite across the $T=0$~K AF-PM transition. Whereas the Curie-Weiss law is generally valid for well-localized spins, the self-consistent-renormalization theory for itinerant magnetism, which extends Stoner theory, shows that correlated spin fluctuations can drive Curie-Weiss like behavior at high $T$ \cite{Santiago_2017, Moriya_1985,Takahashi_2013}.  As shown in the SM, the Rhodes-Wohlfarth ratio \cite{Rhodes_1963} calculated from our data is $1.5$ to $3$ which indicates itinerant magnetism \cite{SM}.  The SM also presents an analysis using Takahashi's theory for itinerant magnets \cite{SM}.

We next relate $\chi$ to the electronic structure by plotting the partial $\rho(E)$ contributed by the Co and Fe orbitals for $x=0$, $0.125$, and $0.25$ and $y=0$ in the inset of Fig.~\ref{Fig:chiave_allx}.  The total and partial $\rho(E)$ for $x=0$ and $y=0$ are given in the SM \cite{SM}. A large peak crosses $E_{\text{F}}$ which has contributions from a flat band with Co $d_{x^2-y^2}$ orbital character. The flat band's density of states drives Stoner FM when $\alpha_0> 1$.  This is supported by work showing that the absence of magnetic order in $A$Co$_2$As$_2$, $A =$~Sr and Ba, is a consequence of the flat band lying above $E_{\text{F}}$ \cite{Mao_2018}.  Our DFT calculations indicate an almost rigid shift in $E_{\text{F}}$ with increasing hole doping $x$ with some broadening of the peak in $\rho(E)$ due to the disorder introduced by substituting Fe for Co.  Thus, increasing $x$ pushes $E_{\text{F}}$ below the flat band and decreases $\rho(E_{\text{F}})$. Taken together, our $\chi(T,x)$ and DFT results point to a Stoner-type transition where $x$ tunes $\alpha_0$. When $\alpha_0<1$, $\mu$ vanishes and the continued decrease in $\mu_{\text{eff}}$ with increasing $x$ indicates that $\mu_{\text{fluct}}$ is also  strongly suppressed.  INS can verify this hypothesis by measuring the spin fluctuations  throughout the Brillouin zone. 

\begin{figure}[]
	\centering
	\includegraphics[width=1.0\linewidth]{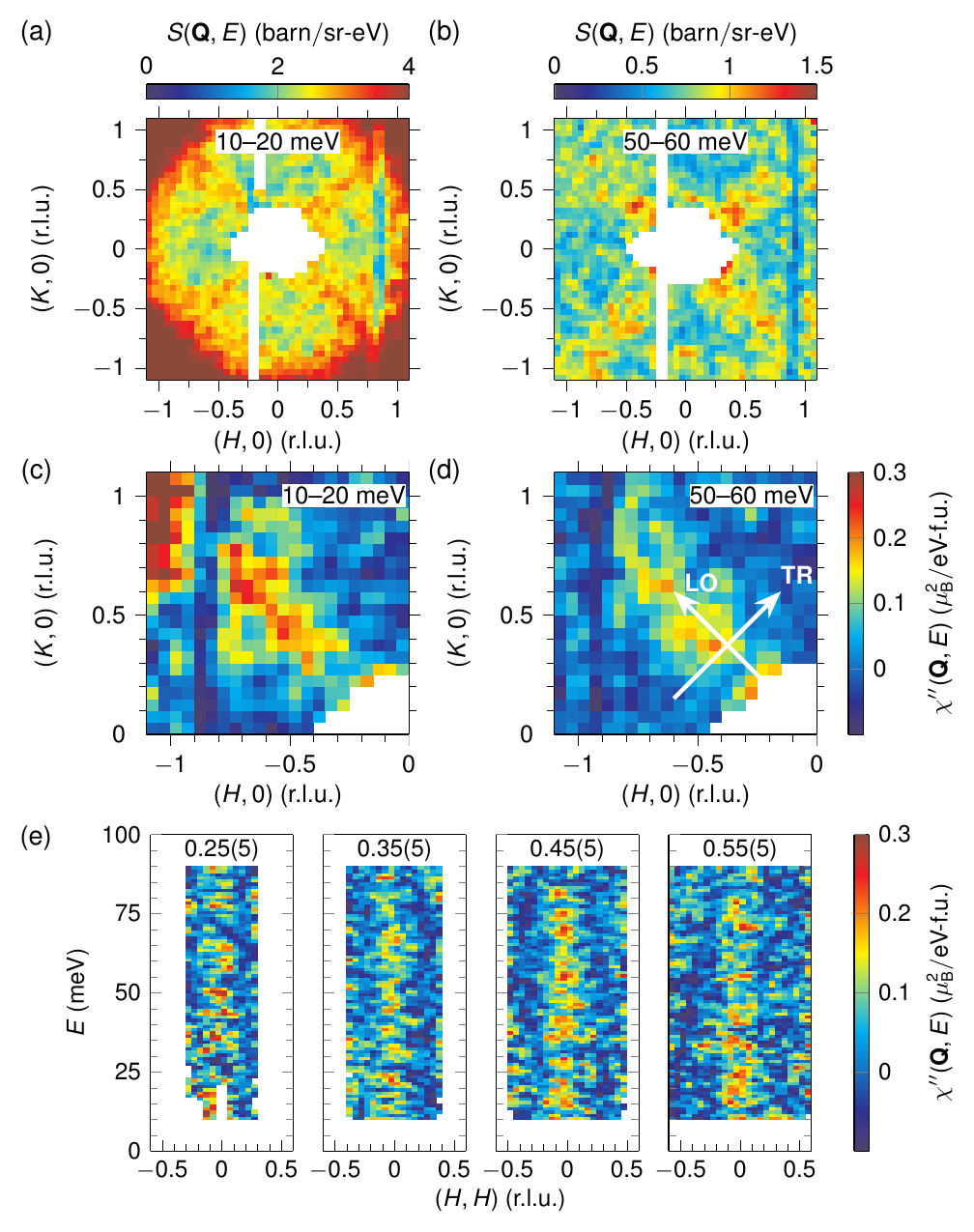}
	\caption{  \label{Fig:Slices} (a),(b) Slices of the INS cross section $S(\mathbf{Q},E)$ in the $(HK)$ plane at $T=5.5$~K integrated over (a) $E=10$ to $20$~meV and (b) $50$ to $60$~meV. (c),(d)  Data corresponding to (a) and (b), respectively, plotted as $\chi^{\prime\prime}(\mathbf{Q},E)$ after an isotropic background subtraction \cite{SM} and averaging over symmetry-equivalent quadrants of the $(HK)$ plane.  Data in (a) and (c) [(b) and (d)] are for $E_{\text{i}}=75$~meV ($125$~meV).  The transverse (TR) $[HH]$ and longitudinal (LO) $[-KK]$ directions are indicated in (d).  (e) TR slices of $E_{\text{i}}=125$~meV data corresponding to (d). TR slices for $E_{\text{i}}=75$~meV are shown in the SM \cite{SM}.  From left to right, plots are for integration ranges of $(-K,K)=(-0.25\pm0.05, 0.25\pm0.05)$, $(-0.35\pm0.05, 0.35\pm0.05)$, $(-0.45\pm0.05, 0.45\pm0.05)$, and $(-0.55\pm0.05, 0.55\pm0.05)$~r.l.u. }  
\end{figure}

Constant-energy slices of the INS cross section $S(\mathbf{Q},E)$  in the $(HK)$  plane for $x=0.15$ are presented in Figs.~\ref{Fig:Slices}(a) and \ref{Fig:Slices}(b). Figures~\ref{Fig:Slices}(c) and \ref{Fig:Slices}(d) show the data plotted as the imaginary part of the dynamical magnetic susceptibility $\chi^{\prime\prime}(\mathbf{Q},E)$ after subtracting off an isotropic and nonmagnetic background estimated from the main data set  and averaging over symmetry-equivalent quadrants of the $(HK)$ plane.  (See the SM~\cite{SM} and Ref.~[\onlinecite{Tucker_2015}] for more details.)  The arrows in Fig.~\ref{Fig:Slices}(d) indicate the transverse (TR) $[HH]$ and longitudinal (LO) $[-KK]$ directions.

Similar to data for $x=0$ \cite{Sapkota_2017}, magnetic scattering in Figs.~\ref{Fig:Slices}(a)--\ref{Fig:Slices}(d) extends longitudinally from $(0,0)$ and is much sharper in the TR direction.  Previous INS data for cobalt arsenides demonstrate weak magnetic intensities due to the combination of a small $\mu_{\text{fluct}}$  and a large energy scale \cite{Sapkota_2017,Jayasekara_2013,LiY_2019,Li_2019}.  By normalizing $S(\mathbf{Q},E)$ for $x=0$ and $0.15$ by the mass of the sample used, we find that the magnetic scattering is $100$ times weaker for $x=0.15$ than for $x=0$ and is close to the limit of detection.

\begin{figure}[]
	\centering
	\includegraphics[width=1.0\linewidth]{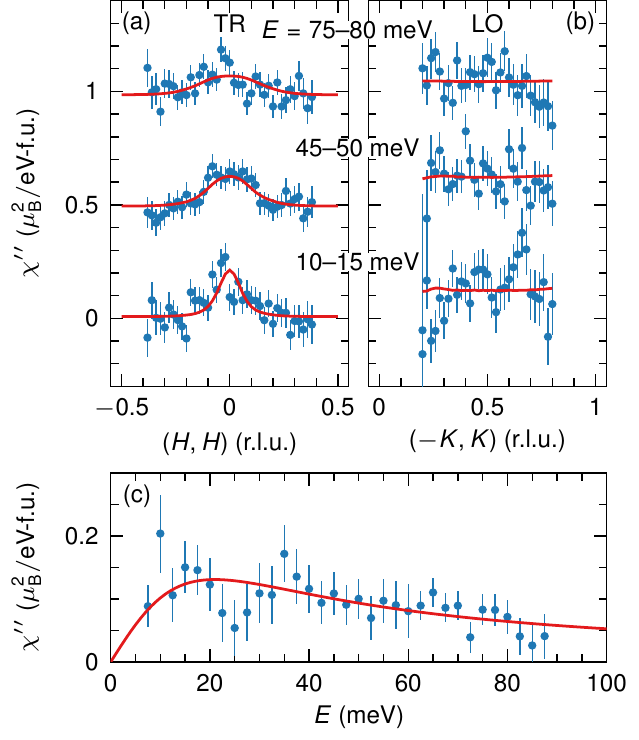}
	\caption{ \label{Fig:cuts} (a) Transverse (TR) (left) and (b) longitudinal (LO) (right) cuts of $\chi^{\prime\prime}(\mathbf{Q},E)$ at  $T=5.5$~K for values of $E$ corresponding to $L\approx1$, $3$, and $5$~r.l.u.  TR (LO) cuts are integrated over $0.2$ to $0.7$~r.l.u.\ ($-0.1$ to $0.1$~r.l.u.)  in the LO (TR) direction.  Datasets are offset by $0.5~\mu_{\text{B}}^2/\text{eV-f.u.}$ Data for $E=10$ to $15$~meV ($E\ge45$~meV ) are for $E_{\text{i}}=75$~meV ($125$~meV). (c) $\chi^{\prime\prime}(E)$ at $T=5.5$~K from integrating over $-0.1$ to $0.1$~r.l.u.\ ($0.2$ to $0.7$~r.l.u.) along the TR (LO) direction.  The $E=7.5$~meV point is from $E_{\text{i}}=75$~meV data, points between $10 $ and $60$~meV are the average of  $E_{\text{i}}=75$ and $125$~meV data, and $E>60$~meV data  correspond to $E_{\text{i}}=125$~meV.   An isotropic background subtraction has been performed \cite{SM} and data were averaged over symmetry-equivalent quadrants of the $(HK)$ plane.  Lines are the results of fits described in the text.}
\end{figure}

Figure~\ref{Fig:Slices}(e) shows $\chi^{\prime\prime}$ in the $E$-$[HH]$ plane for incremental integration ranges along the LO direction.  $\chi^{\prime\prime}(E)$ is steep and extends past $90$~meV, which is characteristic of itinerant magnetism \cite{Moriya_1985}. Figures~\ref{Fig:cuts}(a) and \ref{Fig:cuts}(b) show cuts of $\chi^{\prime\prime}$ for the TR and LO directions, respectively, for different $E$.     The TR width of $\chi^{\prime\prime}$ is only slightly wider than the calculated experimental resolution \cite{SM} and slightly broadens with increasing $E$.  For the LO direction,  $\chi^{\prime\prime}$ is practically constant with increasing $Q$ for a given $E$ and exhibits an overall change in magnitude consistent with the $\chi^{\prime\prime}(E)$ cut in Fig.~\ref{Fig:cuts}(c).

The cut in  Fig.~\ref{Fig:cuts}(c) is for integration ranges of $(H,H)=-0.1$ to $0.1$~r.l.u.\ and $(-K,K)=0.2$ to $0.7$~r.l.u.  $\chi^{\prime\prime}(E)$ peaks around $20$~meV and diminishes with increasing $E$.  The dip at $\approx25$~meV  comes from errors in the background subtraction due to strong contamination by Al phonons.  The lack of periodic variations in $\chi^{\prime\prime}(E)$ indicates practically zero dispersion along $\mathbf{L}$.   Summarizing, other than the much lower intensity, which is consistent with the suppression of $\chi$ in Fig.~\ref{Fig:chiave_allx}, the INS data for $x=0.15$ are similar to those for $x=0$ \cite{Sapkota_2017}, showing quasi-$1$D spin fluctuations.

Magnetic fluctuations in the PM, AF, and superconducting phases of various $122$ pnictides have been described by a diffusive model for $2$D spin fluctuations in a nearly-AF or nearly-FM Fermi-liquid \cite{Moriya_1985} using the $J_1$-$J_2$ Heisenberg model for exchange within the transition metal planes \cite{Inosov_2010,Sapkota_2017, Diallo_2010,Jayasekara_2013, Dai_2015, Sapkota_2018, Sapkota_2017,Tucker_2014,Tucker_2015}.   Within a random-phase approximation, the model gives
\begin{equation}
	\begin{split}
		&\chi^{\prime\prime}(\mathbf{Q},E)=\\*&\frac{\chi^{\prime}(\mathbf{Q}_{\bm{\tau}},0)\Gamma E}{\Gamma^{2}\{1+\frac{4\xi^{2}}{a^{2}}[\eta(c_{+}+c_{-})+c_{+}c_{-}-2\eta-1]\}^2+E^2}\ .
	\end{split}
	\label{Eq:diffusive}
\end{equation}
Here, $\chi^{\prime}(\mathbf{Q}_{\bm{\tau}},0)$ is the staggered static susceptibility at $\mathbf{Q}_{\bm{\tau}}$,  $\xi$ is the magnetic correlation length,  $\Gamma$ quantifies damping of the fluctuations,  $c_{\pm}=\cos{[(q_x\pm q_y)a/2]}$, where $x$ and $y$ denote  perpendicular directions connecting NN spins, and $\mathbf{q}=\mathbf{Q}-\mathbf{Q}_{\bm{\tau}}$.  $\mathbf{Q}_{\bm{\tau}}$ is a reciprocal-lattice position corresponding to the magnetic propagation vector $\bm{\tau}$.  For our case, $\bm{\tau}=(0,0)$. 

We simultaneously fit Eq.~\eqref{Eq:diffusive} to TR, LO, and $E$ cuts of the INS data but found that the value for $\Gamma$ had too much uncertainty.  To mitigate this, we fit the data in Fig.~\ref{Fig:cuts}(c) using the typical quasielastic diffuse magnetic scattering form of $\chi^{\prime\prime}(E)=A E/ (\Gamma^{2}+E^2)$, where $A$ is a scale factor \cite{Shirane_2002}. We next simultaneously fit TR and LO cuts taken every $5$~meV to Eq.~\eqref{Eq:diffusive} while keeping $\Gamma$ fixed.  Lines in Fig.~\ref{Fig:cuts} show examples of the fits with $\chi^{\prime}(\mathbf{Q}_{\bm{\tau}},0)=3.4(3)\times10^{-4}~\mu_{\text{B}}^2/\text{meV-f.u.}$, $\xi/a=1.01(8)$, $\Gamma=21(3)$~meV, and $\eta=-0.97(1)$.  Simulated slices of $\chi^{\prime\prime}(\mathbf{Q},E)$ are shown in the SM \cite{SM}.

With the exception of the extraordinarily small value of $\chi^{\prime}(\mathbf{Q}_{\bm{\tau}},0)$, which is consistent with the $\chi(T)$ data, the determined parameters are analogous to those for $x=0$.  Thus,  our fits find a similar level of frustration exists for $x=0$ and $0.15$ since $\eta\approx-1$ for both compositions.  A table listing the fitted parameters for $x=0$ and $0.15$ and for other $122$ pnictides is given in the SM \cite{SM}.  The INS data can also be used to determine $\mu_{\text{fluct}}$ by integrating $\chi^{\prime\prime}$ over $\mathbf{Q}$ and $E$ \cite{SM}.  We find an extraordinarily small value of $\mu_{\text{fluct}}=0.09(1)~\mu_{\text{B}}/\text{f.u.}$ for $x=0.15$, which is $10$ to $100$ times smaller than $\mu_{\text{fluct}}$ for related compounds \cite{SM}.  

As noted above, even though $\chi^{\prime}(\mathbf{Q}_{\bm{\tau}},0)$ for $x=0$ has yet to be measured on an absolute scale, we know that $S(\mathbf{Q},E)$  is $\approx100$ times stronger for $x=0$ than for $x=0.15$ \cite{Sapkota_2017}.  Thus, $\mu_{\text{fluct}}$ substantially decreases with increasing $x$.  Taken together with the decrease in $\mu_{\text{eff}}$ and the elimination of $\mu$ with increasing $x$, the exceedingly small value of $\mu_{\text{fluct}}$ for $x=0.15$  indicates that hole doping weakens the spin correlations associated with the A-type order. However, since $\eta\approx1$ for both $x=0$ and $0.15$, the weakening is not due to modifying the degree of frustration.  Rather, taking into account the decrease in $\mu$ and $\mu_{\text{fluct}}$ with increasing $x$ and extrapolating the decrease in $\mu_{\text{eff}}^2$ with $x$ in  Fig.~\ref{Fig:structure_J1J2}(d) indicates elimination of the total magnetic moment at $x\approx0.25$.

Since ferromagnetism within the Co planes dominates the magnetic energy scale \cite{Sapkota_2017} and our DFT results indicate that hole doping shifts $E_{\text{F}}$ away from a peak in $\rho(E)$, the quenching of the moment can be explained in terms of a Stoner transition: a decrease in $\rho(E_{\text{F}})$ lowers $\alpha_0=\rho(E_{\text{F}})I$ below $1$ at $x=0.12$ and eliminates the FM order within the Co planes and, in turn, the A-type order.  As evidenced by the further decrease in $\mu_{\text{eff}}$ and $\mu_{\text{fluct}}$, more hole doping eventually completely destroys FM correlations within the Co planes which results in a quenched moment for $x\approx0.25$.

Quenching of the total moment has also been observed for CaFe$_2$As$_2$ which exhibits stripe-AF order.  However, in this case the quenching accompanies a pressure-induced first-order structural phase transition into the collapsed tetragonal (cT) phase characterized by $c/a\lesssim2.8$ \cite{Kreyssig_2008, Goldman_2009,Anand_2012}.  The Fermi surface in the ambient-pressure uncollapsed phase exhibits features consistent with nesting which are not present in the cT phase \cite{Dhaka_2014,Yildrim_2009} and DFT calculations indicate that there disappearance is not due to a rigid shift in $E_{\text{F}}$ \cite{Dhaka_2014}. Ca(Co$_{1-x}$Fe$_{x}$)$_{2-y}$As$_{2}$, on the other hand, crosses over to the cT phase at $x\approx0.5$ \cite{Jayasekara_2017}, well past $x=0.25$.

Finally, a Stoner transition is a quantum phase transition (QPT) since it occurs at $T=0$~K \cite{Brando_2016,Santiago_2017}.  Indeed, the heat capacity data for $x=0.15$ shown in the SM \cite{SM}  indicate that non-Fermi-liquid behavior occurs below $\approx 10$ K which is attributed to a QPT similar to previous reports for Ni$_x$Pd$_{1-x}$ \cite{Nicklas_1999} and YFe$_2$Al$_{10}$ \cite{Wu_2014}.  QPTs in clean itinerant FMs are expected to be first order \cite{Brando_2016}, however, the magnetic transitions in Ca(Co$_{1-x}$Fe$_{x}$)$_{2-y}$As$_{2}$ appear continuous \cite{Jayasekara_2017}. Disorder caused, for example, by Fe substitution and Co vacancies can drive a continuous QPT. On the other hand, even though  the FM Co planes dominate the magnetic energy, A-type AF order is present.  These considerations give compelling reasons to look for quantum fluctuations in other values of $x$, particularly those around $x=0.12$ and $0.25$.

Summarizing, we report the observation of quenching of a magnetic moment by a Stoner-type transition. Our results indicate that in addition to the loss of A-type AF order at $x=0.12$, increasing $x$ eliminates the remaining FM spin correlations in Ca(Co$_{1-x}$Fe$_{x}$)$_{2-y}$As$_{2}$ by $x\approx0.25$ while maintaining extreme frustration.  Our DFT calculations show that increasing $x$ results in hole doping that rigidly shifts $E_{\text{F}}$ away from a peak in $\rho(E)$ from a flat conduction band. Future investigations looking for more evidence of a QPT  for $x$ spanning the disappearance of AF order and the quenching of the moment should be insightful.

\begin{acknowledgments}
	We are grateful for conversations with H.~C.~Walker, A.~I.~Goldman, A.~Kreyssig, P.~P.~Orth,  and D.~Vaknin, and to D. L. Schlagel for assistance with coaligning single crystals.  Work at the Ames Laboratory was supported by the U.~S.~Department of Energy (DOE), Basic Energy Sciences, Division of Materials Sciences \& Engineering, under Contract No.~DE-AC$02$-$07$CH$11358$.  Experiments at the ISIS Neutron and Muon Source were supported by a beamtime allocation RB$1810596$ from the Science and Technology Facilities Council.
\end{acknowledgments}
	

	\newpage
	
	\setcounter{equation}{0}
	\setcounter{figure}{0}
	\setcounter{table}{0}
	\setcounter{page}{1}
	\setcounter{section}{0}
	\makeatletter
	
	\renewcommand{\theequation}{S\arabic{equation}}
	\renewcommand{\thefigure}{S\arabic{figure}}
	\renewcommand{\thetable}{S\arabic{table}}

\onecolumngrid
\begin{center}
	\textbf{{\large Supplemental Material:\\Carrier Tuning of Stoner Ferromagnetism in ThCr$_{\bm{2}}$Si$_{\bm{2}}$-Structure Cobalt Arsenides}}
\end{center}
\vspace{2ex}

\twocolumngrid
\section{The $\bm{J_1}$-$\bm{J_2}$ Heisenberg Model on a Square lattice}

\begin{figure}[]
	\centering
	\includegraphics[width=1\linewidth]{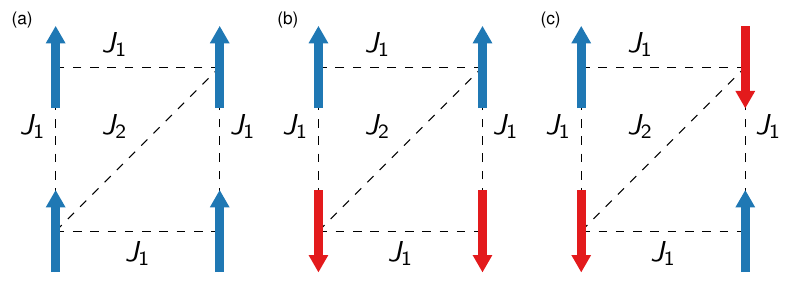}
	\caption{  \label{Fig:J1J2_square} Diagrams for the $J_1$-$J_2$ Heisenberg model on a square lattice.  (a) ferromagnetic (occurring when $-1<\frac{J_1}{2J_2}<0$ with $J_1<0$), (b) stripe-type antiferromagnetic (occurring when $-1<\frac{J_1}{2J_2}<1$ with $J_2>0$), and (c) N\'eel-type antiferromagnetic (occurring when $0<\frac{J_1}{2J_2}<1$ with $J_1>0$) orders are shown.}  
\end{figure}

Local-moment magnets, in which magnetic moments (spins) are due to unpaired electrons localized in orbitals of ions located at specific sites, can be generally described using the Heisenberg model.  The Hamiltonian for the $J_1$-$J_2$ Heisenberg model appropriate for a square lattice is
\begin{equation}
	\mathcal{H}=J_{1}\sum_{\text{NN}}\mathbf{S}_{i}\cdot\mathbf{S}_{j}+J_{2}\sum_{\text{NNN}}\mathbf{S}_{i}\cdot\mathbf{S}_{j}\  , \label{Eq:J1J2Ham}
\end{equation}
where $S_i$ corresponds to a spin at site $i$, $j$ is a site other than $i$, and $J_1$ and $J_2$ are the exchange constants for nearest-neighbor (NN) and next-nearest-neighbor (NNN) interactions, respectively.  Figure~$1$(b) in the main text shows the phase diagram which features ferromagnetic (FM), stripe-type antiferromagnetic (AF), and N\'{e}el-type AF order, and Fig.~\ref{Fig:J1J2_square} shows diagrams of the magnetically order states.  The ground state is tuned by the ratio $\eta=J_1/(2J_2)$.  The borders between states occur at $\eta=1$ and $\eta=-1$ with $J_2>0$, and $\eta=0$ with $J_2<0$.

\section{Stoner Ferromagnetism}
Itinerant magnetism refers to electrons in bands giving rise to the magnetic moment.  Following the descriptions given in Refs.~[\onlinecite{Moriya_1985}] and [\onlinecite{Takahashi_2013}], a general Hamiltonian for an itinerant electron in a crystal can be written as
\begin{align}
	\mathcal{H}&=\mathcal{H}_0+V\label{Eq:Bloch_Ham}\ ,\\ \intertext{where $\mathcal{H}_0$ is the kinetic term and $V$ is the interaction term.   The Hubbard Hamiltonian simplifies this Hamiltonian into}	\mathcal{H}&=\sum_{j,l,\sigma}{t_{jl}a^{\dagger}_{i\sigma}a_{j\sigma}}+U\sum_{j}{n_{j\uparrow}n_{j\downarrow}} ,\label{Eq:HubHam}
\end{align}
where $t_{jl}$ corresponds to hopping of electrons between lattice sites $j$ and $l$, $U$ is the average effective Coulomb repulsion between electrons on the same site, and $\sigma$ stands for the electronic spin which is either up ($\uparrow$) or down ($\downarrow$).   \cite{Moriya_1985}.

The Fourier transform of the kinetic term is
\begin{equation}
	\mathcal{H}_0=\sum_{\mathbf{k}\sigma}\epsilon(\mathbf{k})a^\dagger_{\mathbf{k}\sigma}a_{\mathbf{k}\sigma}\ ,
\end{equation}
where $\mathbf{k}$ is a wavevector and $\epsilon(\mathbf{k})$ is the energy dispersion.   The annihilation and creation operators,  $a$ and $a^\dagger$, respectively, follow the anticommutation relations
\begin{align}
	[a_{\alpha\sigma},a^\dagger_{\beta\sigma^\prime}]_+&=\delta_{\alpha\beta}\delta_{\sigma\sigma^\prime}\ , \\
	[a^\dagger_{\alpha\sigma},a^\dagger_{\beta\sigma^\prime}]_+&=0\ , \\
	[a_{\alpha\sigma},a_{\beta\sigma^\prime}]_+&=0 \ ,
\end{align}
and
\begin{equation}
	n_{\alpha\sigma}=a^\dagger_{\alpha\sigma}a_{\alpha\sigma}\ .
\end{equation}

The Stoner theory for itinerant ferromagnetism is a mean-field approach for considering competition between the kinetic-energy and interaction (electronic-correlation) terms in Eq.~\eqref{Eq:HubHam}.  It considers an electronic-band crossing the Fermi energy $E_{\text{F}}$.  The spontaneous development of a finite magnetization $M$ occurs as the band is split by an energy of $2\Delta$ into two spin-polarized bands in order to lower the total energy.

The two spin-polarized bands have occupations $n_\uparrow$ and $n_\downarrow$, respectively.  The magnetization associated with the splitting is
\begin{equation}
	M=-\frac{1}{2}\sum_{\mathbf{k}}\left<n_{\mathbf{k}\uparrow}-n_{\mathbf{k}\downarrow}\right>=-\frac{N_0}{2}\left<n_\uparrow-n_\downarrow\right>\ ,
\end{equation}
and 
\begin{equation}
	N=\sum_{\mathbf{k}}\left<n_{\mathbf{k}\uparrow}+n_{\mathbf{k}\downarrow}\right>=N_0\left<n_\uparrow+n_\downarrow\right>\\ 
\end{equation}
is the total number of electrons. Here, angular brackets refer to taking the appropriate thermodynamic average, and
\begin{align}
	\left<n_\uparrow\right>&=\frac{N-2M}{2N_0}\\ \intertext{and}
	\left<n_\downarrow\right>&=\frac{N+2M}{2N_0}\ .
\end{align}

Stoner theory uses the Hartree-Fock approximation to handle the interaction term in Eq.~\eqref{Eq:HubHam}:
\begin{align}
	V&=U\sum_{j}{n_{j\uparrow}n_{j\downarrow}}\nonumber\\
	&\approx U\sum_{j}(n_{j\uparrow}\left<n_{\downarrow}\right>+n_{j\downarrow}\left<n_{\uparrow}\right>-\left<n_{\uparrow}\right>\left<n_{\downarrow}\right>)\nonumber\\
	&=U\sum_{\mathbf{k}\sigma}n_{\mathbf{k}\sigma}\left<n_{-\sigma}\right>-N_0U\left<n_{\uparrow}\right>\left<n_{\downarrow}\right>\nonumber\\
	&=I\sum_{\mathbf{k}\sigma} \left(\frac{N}{2}-\sigma M\right)a^\dagger_{\mathbf{k}\sigma}a_{\mathbf{k}\sigma} -I\left(\frac{N^2}{4}-M^2\right)\ , \label{Eq:Hub_HFU}
\end{align}
where
\begin{align}
	N_0I&=U\ ,\\ \intertext{and when $\sigma$ is not printed as a subscript}
	\sigma&=\pm1 \ 
\end{align}
for spin up and spin down, respectively.

Applying a small magnetic field $\mathbf{H}$ that stabilizes a finite $\mathbf{M}$ adds the term \mbox{$\mathcal{H}_h=-\mathbf{H}\cdot\mathbf{M}$} to $\mathcal{H}$.  With the substitution $h=2\mu_{\text{B}}H$:
\begin{align}
	\mathcal{H}_h&=-\mathbf{H}\cdot\mathbf{M}\nonumber\\
	&=-\sum_{\mathbf{k}\sigma}\frac{\sigma h}{2}a^\dagger_{\mathbf{k}\sigma}a_{\mathbf{k}\sigma}\ .
\end{align}
Thus, comparison to Eq.~\eqref{Eq:Hub_HFU} shows that within Hartree-Fock theory the $V$ term in the Hubbard Hamiltonian can be considered as an effective magnetic field with strength $2IM$. Substituting Eq.~\eqref{Eq:Hub_HFU} into Eq.~\eqref{Eq:HubHam} gives
\begin{equation}
	\mathcal{H}=\sum_{\mathbf{k}\sigma}(\epsilon_{\mathbf{k}\sigma}-\mu_{\text{c}})a^\dagger_{\mathbf{k}\sigma}a_{\mathbf{k}\sigma}-I\left(\frac{N^2}{4}-M^2\right)\ ,
\end{equation}
where
\begin{align}
	\epsilon_{\mathbf{k}\sigma}=\epsilon_{\mathbf{k}}+\frac{IN}{2}-\sigma\Delta\\ \intertext{and}
	\Delta=IM+\frac{h}{2}\ .
\end{align}
$\mu_{\text{c}}$ is the chemical potential. 

To find the conditions for FM order, we consider the Landau expansion of the free energy given in terms of magnetization:
\begin{equation}
	F(M,T)=F(0,T)+\frac{a(T)}{2}M^2+\frac{b(T)}{4}M^4+\cdots \ .\label{Eq:F_Land}
\end{equation}
It is shown in Refs.~[\onlinecite{Moriya_1985}] and [\onlinecite{Takahashi_2013}] that
\begin{align}
	a(T)&=\frac{2}{\rho}-2I+\frac{\pi^2}{3}\left(\frac{\rho^{\prime\prime}}{\rho}-\frac{\rho^{\prime2}}{\rho^2}\right)(k_{\text{B}}T)^2+\cdots \label{Eq:a}\\ \intertext{and}
	b(T)&=\frac{1}{\rho^3}\left(\frac{\rho^{\prime2}}{\rho^2}-\frac{\rho^{\prime\prime}}{3\rho}\right)+\cdots \ , \label{Eq:b}
\end{align}
where $^\prime$ ($^{\prime\prime}$) indicates taking the first (second) derivative with respect to $\epsilon$ where $\epsilon\approx E_{\text{F}}$.   FM order occurs for \mbox{$a(0)<0$}.  This leads to the Stoner criteria for FM order:
\begin{equation}
	\alpha_0\equiv I\rho(E_{\text{F}})>1 \ ,
\end{equation}
where $\alpha_0$ is the Stoner parameter.

\section{Details of the Magnetic Susceptibility Measurements}

Magnetization $M$ measurements were made on newly-grown crystals of Ca(Co$_{1-x}$Fe$_{x}$)$_{2-y}$As$_{2}$ with $x=0.036$, $y=0.12(3)$; $x=0.057$, $y=0.12(1)$; $x=0.077$, $y=0.12(3)$; $x=0.12$, $y=0.10(1)$; $x=0.15$, $y=0.00$;  $x=0.19(4)$, $y=0.06(5)$; $x=0.24$, $y=0.00$; and $x=0.35(2)$, $y=0.00(6)$.  Measurements were carried out over a temperature range of $T=1.8$ to $300$~K and under various values of applied magnetic field $H$ using a Quantum Design, Inc., Magnetic Properties Measurement System (MPMS).  The static susceptibility was determined through the relation $\chi\equiv M/H$\@.   The applied fields were $0.1$~T for $x=0$--$0.15$, $1$~T for $x=0.19$ and $0.24$, and $3$~T for $x=0.35$.  Measurements were made for both $\mathbf{H}\parallel\mathbf{c}$ and $\mathbf{H}\perp\mathbf{c}$ to determine $\chi_c$ and $\chi_{ab}$ versus temperature~$T$, respectively.  The spherical (powder) average of $\chi$ was found from $\chi_{\text{ave}}=\frac{2}{3}\chi_{ab}+\frac{1}{3}\chi_{c}$.

\begin{figure*}[]
	\centering
	\includegraphics[width=0.77\linewidth]{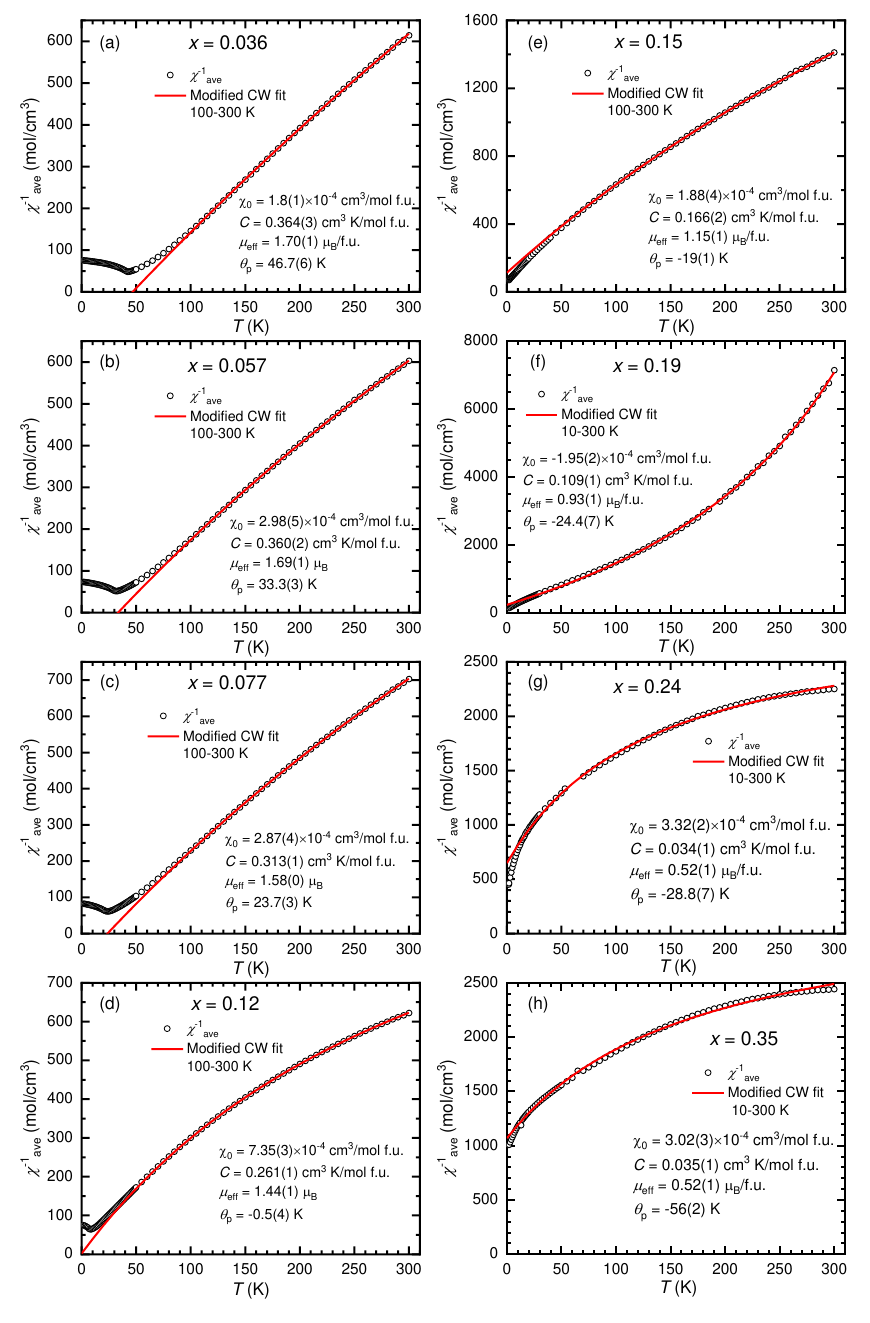}
	\caption{  \label{Fig:chi_inv} Inverse of the powder-averaged magnetic susceptibility $\chi_{\rm ave}^{-1}$ as a function of temperature $T$ for  Ca(Co$_{1-x}$Fe$_{x}$)$_{2-y}$As$_{2}$ single crystals with $x=0.036$, $y=0.12$ (a), $x=0.057$, $y=0.12$ (b), $x=0.077$, $y=0.12$ (c), $x=0.12$, $y=0.10$, (d) $x=0.15$, $y=0.00$ (e),  $x=0.19$, $y=0.06$ (f), $x=0.24$, $y=0.00$ (g), and $x=0.35$, $y=0.00$ (h).  $\chi_{\text{ave}}$ is the spherical average value of $\chi$, as described in the text. The solid curves are fits to Eq.~\eqref{Eq:MCWLaw} where $C$ is the Curie constant, $\theta_{\text{p}}$ is the Weiss temperature, $\chi_0$ is the temperature-independent contribution to the susceptibility, and $\mu_{\text{eff}}$ is the effective moment determined from $C$\@.  The temperature range for each fit is indicated.  }  
\end{figure*}

Figures~\ref{Fig:chi_inv}(a)--\ref{Fig:chi_inv}(h) show  $\chi_{\text{ave}}^{-1}(T)$ data for our crystals.  The red solid lines are fits of the respective data sets by the modified Curie-Weiss Law \cite{Kittel_2004,Ashcroft_2020}
\begin{equation}
	\chi = \chi_0 + \frac{C}{T-\theta_{\text{p}}},
	\label{Eq:MCWLaw}
\end{equation}
where $\chi_0$ is the temperature-independent contribution, $C$ is the Curie constant, and $\theta_{\text{p}}$ is the Weiss temperature.   Here,
\begin{equation}
	C=\frac{N_{\text{A}}\mu^2_{\text{eff}}}{3k_{\text{B}}}\ ,
	\label{Eq:CurieC}
\end{equation}
where $N_{\text{A}}$ is Avogadro's number, $k_{\text{B}}$ is the Boltzmann constant, and $\mu_{\text{eff}}=\sqrt{8C}$ in cgs units is the effective moment per formula unit (f.u.)\ in units of $\mu_{\text{B}}$.

\begin{figure}[]
	\centering
	\includegraphics[width=1\linewidth]{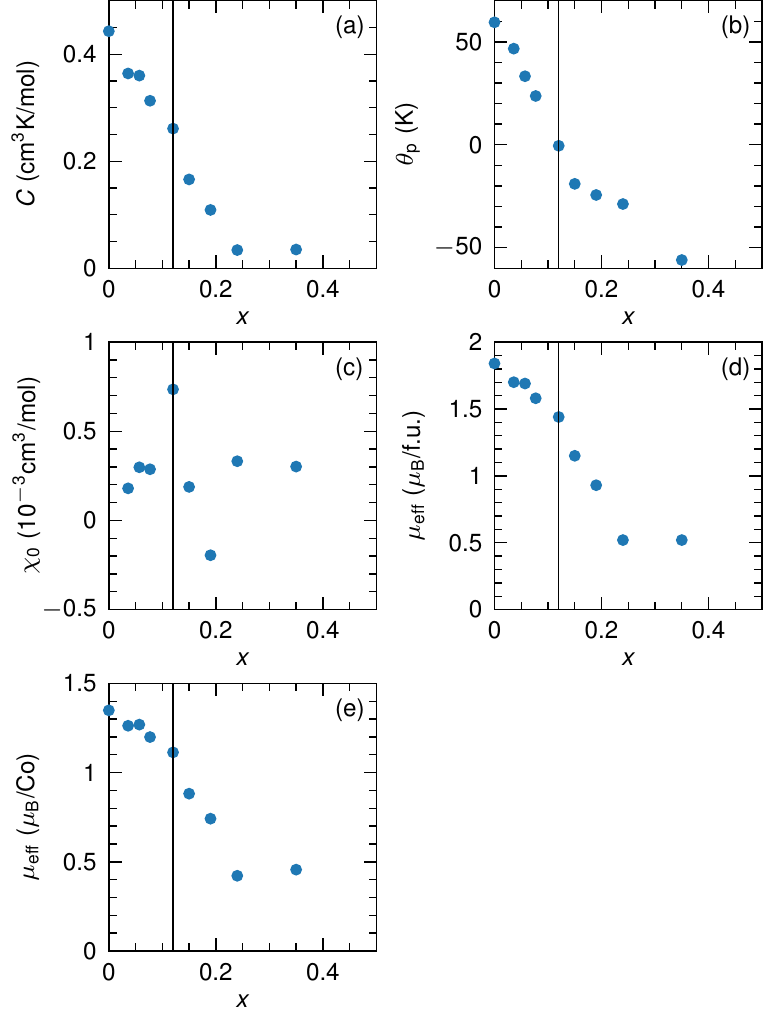}
	\caption{  \label{Fig:chi_pars} Parameters obtained from fits of the data in Fig.~\ref{Fig:chi_inv} by the modified Curie-Weiss Law  [Eq.~\eqref{Eq:MCWLaw}]: The Curie constant $C$~(a), Weiss temperature $\theta_{\rm p}$~(b), temperature-independent susceptibility $\chi_0$~(c), and the effective magnetic moment $\mu_{\rm eff} = \sqrt{8C}$~(d).  Data for $x=0$ are from Ref.~[\onlinecite{Anand_2014}].  (e) $\mu_{\rm eff}$ determined after normalizing $C$ to units of $\text{cm}^3\text{K}/\text{Co}$. The vertical lines at $x=0.12$ indicate the $T=0$~K antiferromagnetic-paramagnetic phase transition. Error bars are smaller than the data markers.}
\end{figure}

\mbox{Figure~\ref{Fig:chi_pars}} shows plots of the fitted parameters versus $x$.  Data for $x=0$ are from Ref.~[\onlinecite{Anand_2014}].  As discussed in the main text, $\mu_{\text{eff}}$ decreases with increasing $x$.   $\theta_{\text{p}}$ also decreases with $x$, crossing zero at the $T=0$~K antiferromagnetic (AF) to paramagnetic (PM) transition at $x=0.12(1)$.  The parameters $C$ and $\theta_{\rm p}$ vary monotonically with composition~$x$, however, $\chi_0$ shows outliers at $x=0.12$ and $0.19$.  At present, the reason for these nonsystematic values is unclear.  Due to the small overall value of $\chi$ and the shapes of $\chi^{-1}(T)$ for $x$ in the quenched moment region ($x\gtrsim0.25$), the fitted parameters for $x=0.24$ and $0.35$ are unreliable.  Figure~\ref{Fig:chi_pars}(e) plots $\mu_{\text{eff}}$ after renormalizing $C$ to units of $\text{cm}^3\text{K}/\text{Co}$.  If we presume a localized-moment picture where substituting Fe for Co causes nonmagnetic dilution, then one may expect that $\mu_{\text{eff}}(x)$ normalized per Co ion would remain constant with increasing $x$.  Figure~\ref{Fig:chi_pars}(e) shows that this is not the case for  Ca(Co$_{1-x}$Fe$_{x}$)$_{2-y}$As$_{2}$.

\begin{figure}[]
	\centering
	\includegraphics[width=1\linewidth]{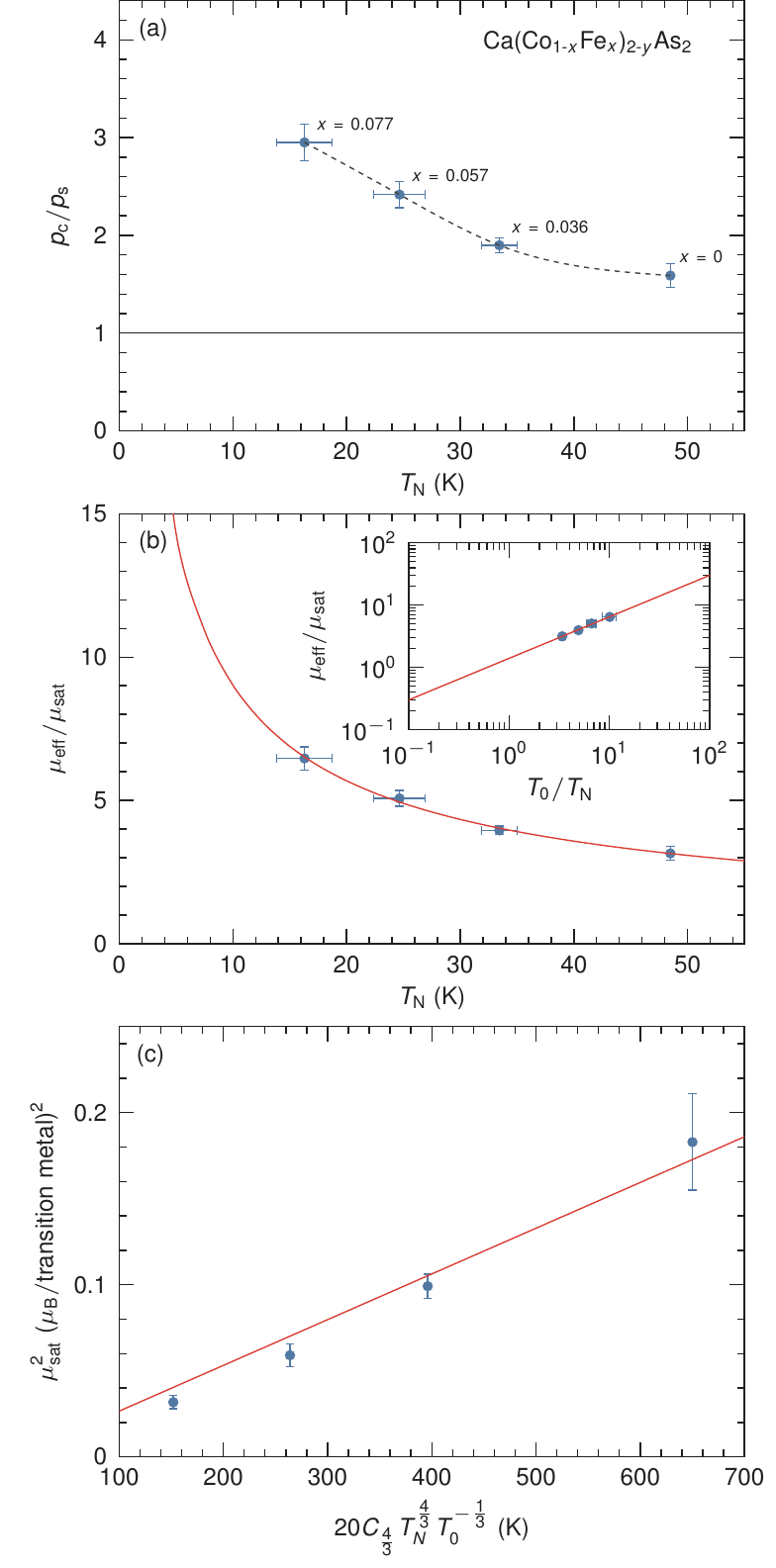}
	\caption{  \label{Fig:RW_plot} (a) Rhodes-Wohlfarth ratio $p_{\text{c}}/p_{\text{s}}$ versus the N\'{e}el temperature $T_{\text{N}}$ for Ca(Co$_{1-x}$Fe$_{x}$)$_{2-y}$As$_{2}$ single crystals. (b) $\mu_{\text{eff}}/\mu_{\text{sat}}$ versus $T_{\text{N}}$.  The curve is a fit to Takahashi's theory, as described in the text. The inset shows the data and results of the fit using logarithmic axes with $T_{0}=164(7)$~K. (c) Plot to determine $T_{\text{A}}$, as described in the text, using $T_{0}=164$~K.  The line shows a linear fit to the data.}
\end{figure}

Next, we use the $\chi$ data  and the Rhodes-Wohlfarth ratio \cite{Rhodes_1963} to discuss the classification of Ca(Co$_{1-x}$Fe$_{x}$)$_{2-y}$As$_{2}$ as itinerant magnets.  We define $p_{\text{c}}=\sqrt{1+\mu_{\text{eff}}^2}-1$ and $p_{\text{s}}=\mu_{\text{sat}}$, where  $\mu_{\text{eff}}$ and $\mu_{\text{sat}}$ (the saturated magnetic moment) are given in units of $\mu_{\text{B}}/\text{transition metal}$.  $\mu_{\text{sat}}$ is equal to the ordered moment per transition metal found from neutron diffraction in the limit of $T\rightarrow0$~K \cite{Takahashi_2013,Santiago_2017,Sangeetha_2019}.   $p_{\text{c}}/p_{\text{s}}=1$ for local-moment magnets and $p_{\text{c}}/p_{\text{s}}>1$ for  itinerant magnets \cite{Rhodes_1963}.  

Figure~\ref{Fig:RW_plot}(a) plots $p_{\text{c}}/p_{\text{s}}$ versus $T_{\text{N}}$, where we have used the equations given in Ref.~[\onlinecite{Jayasekara_2017}] to determine $T_{\text{N}}$ and $\mu$. These values of $\mu$ are for either $T=4$ or $1.8$~K (i.e.\ not for $T=0$~K), but the magnetic order parameter plots in Ref.~[\onlinecite{Jayasekara_2017}] indicate that this should introduce negligible error. Figure~\ref{Fig:RW_plot}(a) shows that all of the compounds have a ratio of $p_{\text{c}}/p_{\text{s}}>1$ and $p_{\text{c}}/p_{\text{s}}$ increases with increasing $x$.   Some limits on using $p_{\text{c}}/p_{\text{s}}$ as a metric are discussed in Ref.~[\onlinecite{Rhodes_1963}].

Takahashi gives a theory that goes beyond the Stoner and  Self-Consistent-Renormalization (SCR) theories \cite{Moriya_1985} by considering that the coefficient $b$ in Eq.~\eqref{Eq:F_Land} is temperature dependent \cite{Takahashi_2013}.   Takahashi explains that only considering the temperature dependence of the coefficient $a$  in Eq.~\eqref{Eq:F_Land}, as done for the Stoner and SCR theories, results in $M$ discontinuously going to zero at the Curie temperature $T_{\text{C}}$ when entering into the paramagnetic phase.

A modified version of the Rhodes-Wohlfarth ratio based on the theory of Takahashi is given in Chapter~$3$ of Ref.~[\onlinecite{Takahashi_2013}] and allows for direct comparison between results from inelastic neutron scattering (INS) and magnetic susceptibility data. We first define Takahashi's quantity $T_{0}$ via $2k_{\text{B}}T_0=\Gamma_{q_{\text{B}}}$, which is a measure of the inverse of the lifetime of spin fluctuations at the zone-boundary wave vector $q_\text{B}$  \cite{Takahashi_2013}.  Following Ref.~[\onlinecite{Takahashi_2013}], we take $\Gamma_{q_{\text{B}}}\approx\Gamma q_{\text{B}}^3$ and use $q_{\text{B}}=\pi\sqrt{2}/a$ and the values for $\Gamma$  given in Table~\ref{Tab:fit_params}. We find $T_0=174\pm25$~K for $x=0.15$ and $T_0=166\pm33$~K for $x=0$.

Next, Takahashi's modification of the RW ratio for an itinerant FM is \cite{Takahashi_2013}
\begin{equation}
	\frac{\mu_{\text{eff}}}{\mu_{\text{sat}}}\approx1.4\left(\frac{T_{0}}{T_{\text{C}}}\right)^{\frac{2}{3}} \ . \label{Eq:Taka}
\end{equation}
We test this relation by using the same values of $\mu_{\text{eff}}$ and $\mu_{\text{sat}}$ used above for the Rhodes-Wohlfarth ratio, and, as above, use $T_{\textrm{N}}$ instead of $T_{\text{C}}$.  A plot of $\mu_{\text{eff}}/\mu_{\text{sat}}$ versus $T_{\text{N}}$ is given in Fig.~\ref{Fig:RW_plot}(b) where the solid line shows a fit to the data using Eq.~\eqref{Eq:Taka}.  The fit yields  $T_{0}=164(7)$~K which is consistent with the values obtained from the INS data.  

One also can find Takahashi's quantity $T_{\text{A}}$ which is a measure of the spectral dispersion in momentum space \cite{Takahashi_2013}.  For ferromagnets, $T_{\text{A}}$ is related to $T_0$ and $\mu_{\text{sat}}$ through
\begin{equation}
	\mu_{\text{sat}}^2=\frac{20}{T_{\text{A}}}C_{\frac{4}{3}}T_{\text{C}}^{\frac{4}{3}}T_0^{-\frac{1}{3}} \ ,
\end{equation}
where $C_{4/3}=1.006089\ldots$\ and we have used a spectroscopic splitting factor (g-factor) of $g=2$.  Figure~\ref{Fig:RW_plot}(c) plots our data in terms of this equation with the value of $T_0=164$~K found from our susceptibility and previous neutron diffraction data. We again use $T_{\text{N}}$ in place of $T_{\text{C}}$.  The line shows a linear fit which determines that $T_{\text{A}}=3763\pm583$~K.  The values of $T_0$ and $T_{\text{A}}$ determined from our data are consistent with results for other itinerant ferromagnets given in Chapter~$3$ of Ref.~\onlinecite{Takahashi_2013}.

Equation~\eqref{Eq:Taka} allows one to predict $T_0$ from the values of $\mu_{\text{eff}}$, $\mu_{\text{sat}}$, and $T_{\text{C}}$ found from magnetic susceptibility data for an itinerant ferromagnet.  The good agreement between $T_0$ determined for Ca(Co$_{1-x}$Fe$_{x}$)$_{2-y}$As$_{2}$ from INS data and $T_0$ determined from our magnetic susceptibility and previous neutron diffraction data  again shows that itinerant ferromagnetism within the Co planes dominates the magnetic energy scale.

$T_0$ and $T_{\text{A}}$ can be used to calculate the ground-state magnetic isotherm from
\begin{equation}
	\begin{split}
		F&=F(T=0)+\frac{1}{2}a(T=0)M^2+\frac{1}{4}b(T=0)M^4\\
		&=F(T=0)+\frac{1}{2g^2\mu_{\text{B}}^2\chi}M^2+\frac{F_1}{4g^4\mu_{\text{B}}^4N_0^3}M^4\ .
	\end{split}
\end{equation}
Here, $F_1=(2T_{\text{A}}^2/(15cT_0)$ and $c=1/2$ for a Lorentzian distribution of the zero-point (quantum) spin fluctuations. The absence of thermal fluctuations in the ground state means that quantum spin fluctuations give rise to $b$.   From our INS data, we find that $T_0$ may increase between $x=0$ and $x=0.15$ which would be consistent with the quantum fluctuations indicated by the heat capacity data for $x=0.15$ presented below.  On the other hand, the decrease in $\mu$ and $\mu_{\text{fluct}}$ with increasing $x$ is consistent with the Stoner and SCR theories where $b$ is only determined by the density of states at $E_{\text{F}}$ and its first and second derivatives with respect to $E$ at $E_{\text{F}}$ [Eq.~\eqref{Eq:b}].  Further studies examining the presence and nature of quantum fluctuations with increasing $x$ should prove insightful.

\section{Details of the Inelastic Neutron Scattering Experiments}
Inelastic neutron scattering (INS) experiments were performed on the MERLIN spectrometer at the ISIS Neutron and Muon Source at the Rutherford Appleton Laboratory \cite{MERLIN_2018}.  Seven crystals ($2.1$~g total) of Ca(Co$_{0.85}$Fe$_{0.15}$)$_{2}$As$_{2}$ were coaligned on Al plates for INS experiments with the $(H0L)$ reciprocal-lattice plane horizontal. Data were collected as functions of energy $E$ and neutron-momentum transfer $\mathbf{Q}$.  The sample was mounted in a He closed-cycle refrigerator and cooled to a temperature of $T=5.5$~K.  Incident neutron energies of $E_{\text{i}}=125$ and $75$~meV were used with chopper rotation rates of $f=250$ and $350$~Hz, respectively.  The sample's $\mathbf{c}$ axis was fixed parallel to the incident neutron beam, which links $L$ to $E$.  The lattice constants at $T=5.5$~K  are  $a=3.944(1)$~\AA\ and $c=10.35(1)$~\AA, which agree with values reported previously \cite{Jayasekara_2017}.   Measurements of a vanadium standard have been used to put the INS cross section into absolute units.  The resolution in the $[HH]$ direction at the $(0.5,0.5,1)$ reciprocal-lattice position is calculated to be $\approx0.04$~r.l.u.$=0.09$~\AA$^{-1}$ for $E=12.5$~meV. 

\begin{figure}[]
	\centering
	\includegraphics[width=1.0\linewidth]{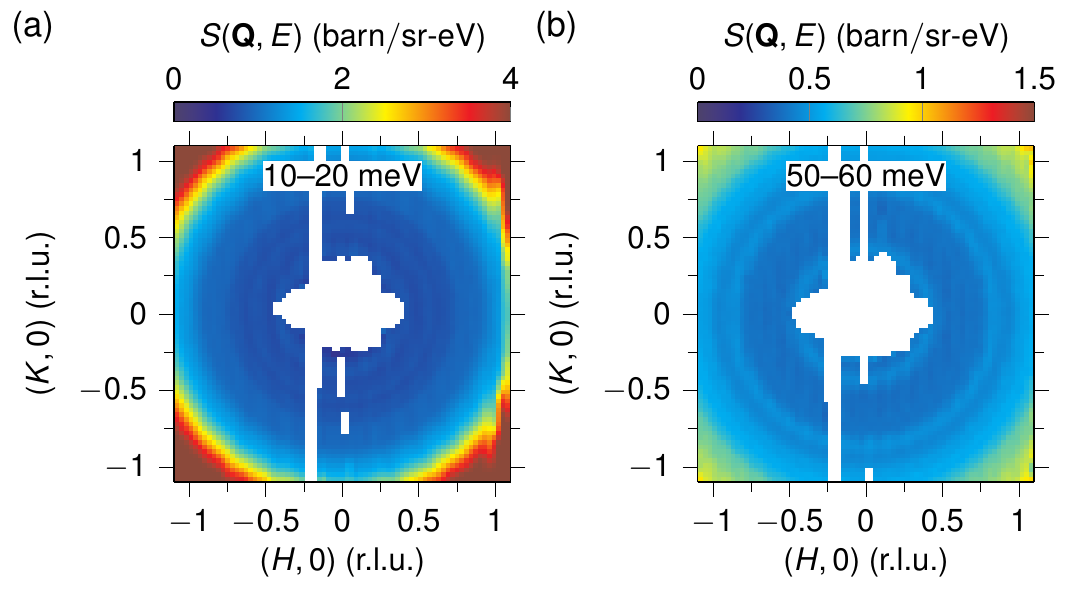}
	\caption{  \label{Fig:bckd_slices} Background inelastic neutron scattering cross sections corresponding to Figs.~$3$(a) and $3$(b) of the main text calculated by the technique discussed in the SM.  Panel (a) is for $E_{\text{i}}=75$~meV and an integration range of $E=10$--$20$~meV.  Panel (b) is for $E_{\text{i}}=125$~meV and an integration range of $E=50$--$60$~meV.   }  
\end{figure}

\begin{figure}[]
	\centering
	\includegraphics[width=1.0\linewidth]{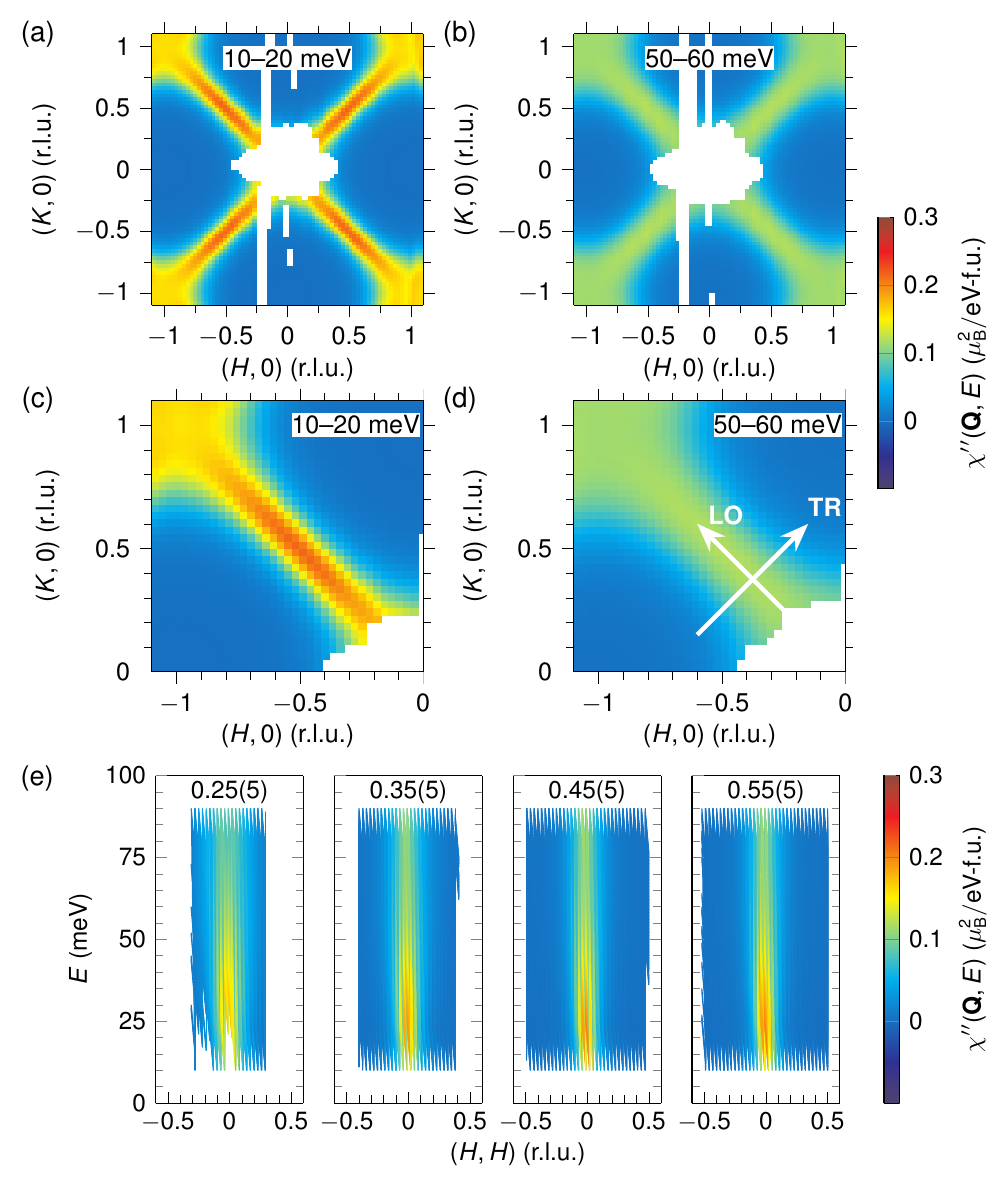}
	\caption{  \label{Fig:Sim_slices} (a),(b) Simulations of $E$ slices of the imaginary part of the magnetic susceptibility $\chi^{\prime\prime}$ in the $(HK)$ plane calculated using the diffusive model [Eq~($1$) in the main text] with the fitted parameters for integration ranges of (a) $E=10$ to $20$~meV and (b) $50$ to $60$~meV. (c),(d)  Data corresponding to (a) and (b), respectively, after averaging over symmetry-equivalent quadrants of the $(HK)$ plane. The transverse (TR) $[HH]$ and longitudinal (LO) $[-KK]$ directions are indicated in (d).  (e) Simulations of TR slices using the diffusive model [Eq~($1$) in the main text] with the fitted parameters.  From left to right, plots are for integration ranges of $(-K,K)=(-0.25\pm0.05, 0.25\pm0.05)$, $(-0.35\pm0.05, 0.35\pm0.05)$, $(-0.45\pm0.05, 0.45\pm0.05)$, and $(-0.55\pm0.05, 0.55\pm0.05)$~r.l.u.  }  
\end{figure}

In order to better characterize the observed weak magnetic scattering, we estimated and subtracted off contributions from the polycrystalline Al sample holder.  Ideally such background scattering is isotropic in $\mathbf{Q}$ due to powder averaging. So, we averaged over the INS cross section $S(\mathbf{Q},E)$ using thin shells in $(|\mathbf{Q}|,E)$ to obtain an estimate of the nonmagnetic background cross section  $S_{\text{bg}}(|\mathbf{Q}|,E)$.  For the final data analysis, we also applied a mask over the magnetic INS to improve $S_{\text{bg}}(|\mathbf{Q}|,E)$.  This isotropic function is then remapped into the full $(\mathbf{Q},E)$ space and subtracted from the full data set.   Examples of the calculated $S_{\text{bg}}(|\mathbf{Q}|,E)$ corresponding to Figs.~$3$(a) and $3$(b) of the main text are shown in Fig.~\ref{Fig:bckd_slices}. Simulations of $\chi^{\prime\prime}(\mathbf{Q},E)$ using the fitted parameters from fits to the diffusive model discussed in the main text were also performed.  Slices of the simulated $\chi^{\prime\prime}(\mathbf{Q},E)$ corresponding to those in  Figs.~$3$(c)--$3$(e) are shown in Fig.~\ref{Fig:Sim_slices}.

\begin{figure}[]
	\centering
	\includegraphics[width=1.0\linewidth]{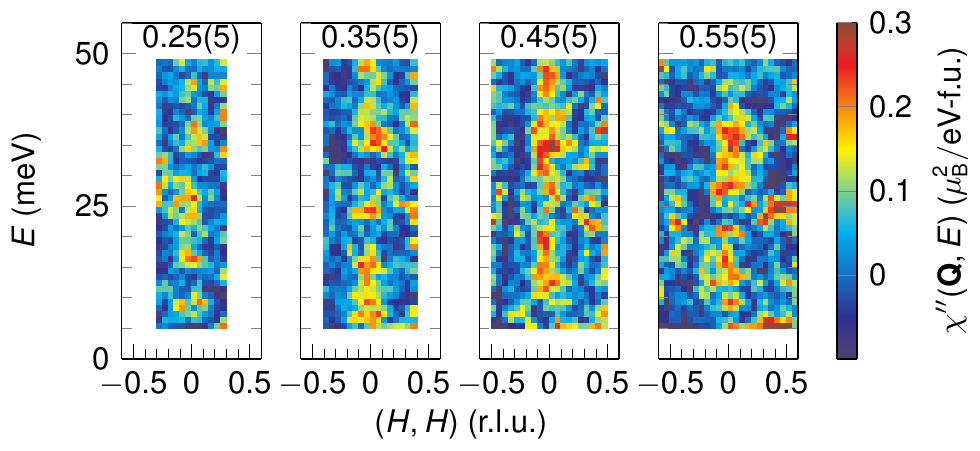}
	\caption{  \label{Fig:TR_Slices_Co1} Transverse slices of the the imaginary part of the magnetic susceptibility $\chi^{\prime\prime}$ from $E_{\text{i}}=75$~meV data which correspond to data in Fig.~$3$(c) of the main text.  From left to right, plots are for integration ranges of $(-K,K)=(-0.25\pm0.05, 0.25\pm0.05)$, $(-0.35\pm0.05, 0.35\pm0.05)$, $(-0.45\pm0.05, 0.45\pm0.05)$, and $(-0.55\pm0.05, 0.55\pm0.05)$~r.l.u.  }  
\end{figure}

$\chi^{\prime\prime}$ is related to $S$ by
\begin{equation}
	\begin{split}
		\chi^{\prime\prime}&(\mathbf{Q},E)\\ &=\frac{2\pi}{(\gamma r_0)^2}\frac{S(\mathbf{Q},E)-S_{\text{bg}}(Q,E)}{f^2(Q)}(1-e^{-E/k_\text{B}T}) \ ,
	\end{split}
	\label{Eq:imchi}
\end{equation}
where $(\gamma r_0)^2=0.2906$~barn/sr, $k_{\text{B}}$ is the Boltzmann constant, and $f(Q)$ is the neutron magnetic form factor for Co$^{1+}$.  The choice of Co$^{1+}$ is discussed below.  For completeness and comparison to Fig.~$3$(e) of the main text, Fig.~\ref{Fig:TR_Slices_Co1} shows slices of $\chi^{\prime\prime}$ versus $E$  along the transverse ($[HH]$ direction) for $E_{\text{i}}=75$~meV.  

\begin{table*}
	\caption{ \label{Tab:fit_params} Fit parameters for the diffusive model described in the main text or a corresponding model including exchange along $\mathbf{c}$.  Also given are the N\'eel  temperature $T_{\text{N}}$ and superconducting transition temperature $T_{\text{c}}$, if antiferromagnetic or superconducting transitions occur. $T$ is the temperature for the fitted data.  The column labeled structure indicates the corresponding tetragonal phase of the compounds either at the measurement temperature or at a temperature just above the orthorhombic structural transition temperature, if an orthorhombic phase occurs.  cT stands for collapsed tetragonal and Tet stands for the uncollapsed-tetragonal phase.}
	\begin{ruledtabular}
		\begin{tabular}{ld{1.4}d{2.4}d{1.4}d{1.7}d{4.3}d{0}d{-1}d{-1}c}	
			\multicolumn{1}{c}{}&\multicolumn{1}{c}{$\sqrt{\mu_{\text{fluct}}^2}$}&\multicolumn{1}{c}{$\eta$}&\multicolumn{1}{c}{$\xi/a$}&\multicolumn{1}{c}{$\chi^{\prime}(\mathbf{Q}_{\bm{\tau}},0)$}&\multicolumn{1}{c}{$\Gamma$} &\multicolumn{1}{c}{$T_{\text{N}}$}&\multicolumn{1}{c}{$T_{\text{c}}$}& \multicolumn{1}{c}{$T$}&Structure\\
			\multicolumn{1}{c}{}&\multicolumn{1}{c}{$(\mu_{\text{B}}/$f.u.$)$}&\multicolumn{1}{c}{}&\multicolumn{1}{c}{}&\multicolumn{1}{c}{$(\mu_{\text{B}}^2/\text{meV-fu})$}&\multicolumn{1}{c}{$($meV$)$} &\multicolumn{1}{c}{$($K$)$}&\multicolumn{1}{c}{$($K$)$}& \multicolumn{1}{c}{$($K$)$}& \\
			\hline
			Ca(Co$_{0.85}$Fe$_{0.15}$)$_{2}$As$_{2}$& 0.09(1)& -0.97(1)& 1.01(8)&  0.00034(3)& 21(3)& \text{---}& \text{---}&5.5& cT\\
			CaCo$_{1.86}$As$_{2}$~\footnote[1]{Reference~\cite{Sapkota_2017}.}& \multicolumn{1}{c}{---}& -1.03(2)& 1.0(4)&\multicolumn{1}{c}{---}& 20(4)&   52(1)& \text{---}& 8& cT\\
			SrCo$_{2}$As$_{2}$~\footnote[2]{Reference~\cite{Li_2019}.}& 0.38(3)&-0.48(4)& 1.93(7)& 0.101(4)& 7.6(4)&   \text{---}& \text{---}& 5& Tet\\
			Ca(Co$_{0.03}$Fe$_{0.97}$)$_{2}$As$_{2}$~\footnote[3]{Reference~\cite{Sapkota_2018}.}& 0.89(3)&0.45(7)& 2.1(8)&  0.15& 9.8(5)&  \text{---}& 14.0(5)& 90& Tet\\
			Ca(Co$_{0.026}$Fe$_{0.974}$)$_{2}$As$_{2}$~\footnotemark[3]& 0.85(3)& 0.44(7)&  2.30(5)&0.14(1)& 7.8(5)&  64(8)& \text{---}& 90& Tet\\
			CaFe$_{2}$As$_{2}$~\footnote[4]{Reference~\cite{Diallo_2010}}&1.0(1)& 0.3(2)&  2.0(4)&  0.20(5)& 10& 172& \text{---}&180& Tet\\
			Ba(Co$_{0.040}$Fe$_{0.960}$)$_{2}$As$_{2}$~\footnote[5]{References~\cite{Tucker_2014} and \cite{Tucker_2015}.}& 1.00(14)& 0.57&2.6(2)& 1.4(3)& 10.4(6)&   65& 9& 20& Tet\\
			Ba(Co$_{0.047}$Fe$_{0.953}$)$_{2}$As$_{2}$~\footnotemark[5]& 1.04(6) &0.57&2.91(8)&  1.61(13)& 8.9(4)&  51& 15& 25& Tet\\
			Ba(Co$_{0.055}$Fe$_{0.945}$)$_{2}$As$_{2}$~\footnotemark[5]&  0.91(8)& 0.57&2.3(1)& 0.82(12)& 7.9(4)&  37& 21& 30& Tet\\
		\end{tabular}
	\end{ruledtabular}
\end{table*}

We can use the INS data to find the fluctuating magnetic moment $\mu_{\text{fluct}}$ via the relation
\begin{equation}
	\mu_{\text{fluct}}^2=\frac{3}{\pi}\frac{\int\chi^{\prime\prime}(\mathbf{Q},E)(1-e^{-E/k_\text{B}T})^{-1}\text d\mathbf{Q}\text dE}{\int\text d\mathbf{Q}}\ ,
	\label{Eq:fluc_mom}
\end{equation}
where the integral over $\mathbf{Q}$ is taken over the Brillouin zone and we use a range of $10$ to $200$~meV for the integral over $E$.   We determined $\mu_{\text{fluct}}$ by inserting Eq.~($1$) of the main text into the numerator and using the fitted parameters from fits to the diffusive model described in the main text.  The choice of $E=200$~meV as the upper bound of the integration is an overestimate of the spectral cutoff based on previous analysis of the spectra of CaCo$_{2-y}$As$_2$.  It sets a conservative upper limit to the size of $\mu^2_{\text{fluct}}$.  Table~\ref{Tab:fit_params} lists $\mu_{\text{fluct}}$ and fitted parameters  for various $122$ compounds \cite{Sapkota_2017,Sapkota_2018,Li_2019,Diallo_2010,Tucker_2014,Tucker_2015}.  The much weaker value of $\mu_{\text{fluct}}$ for Ca(Co$_{0.85}$Fe$_{0.15}$)$_{2}$As$_{2}$ is apparent, as well as the variance of $\eta=J_1/(2J_2)$ and the magnetic correlation length $\xi$ between different compounds.

We now discuss the choice of a Co$^{1+}$ form factor.  It has been proposed that the collapsed-tetragonal (cT) phase of CaCo$_{1.86}$As$_2$ has Co$^{1+}$ whereas Co$^{2+}$ is present in the (uncollapsed) tetragonal phase \cite{Quirinale_2013}.   The inferred valence is based on studies of many body-centered-tetragonal ThCr$_{2}$Si$_{2}$-structure compounds \cite{Hoffmann_1985,Reehuis_1990,Reehuis_1998}. Since Ca(Co$_{0.85}$Fe$_{0.15}$)$_{2}$As$_{2}$ exists in the cT phase, we use $f(Q)$ for Co$^{1+}$.

\begin{figure}[]
	\centering
	\includegraphics[width=1.0\linewidth]{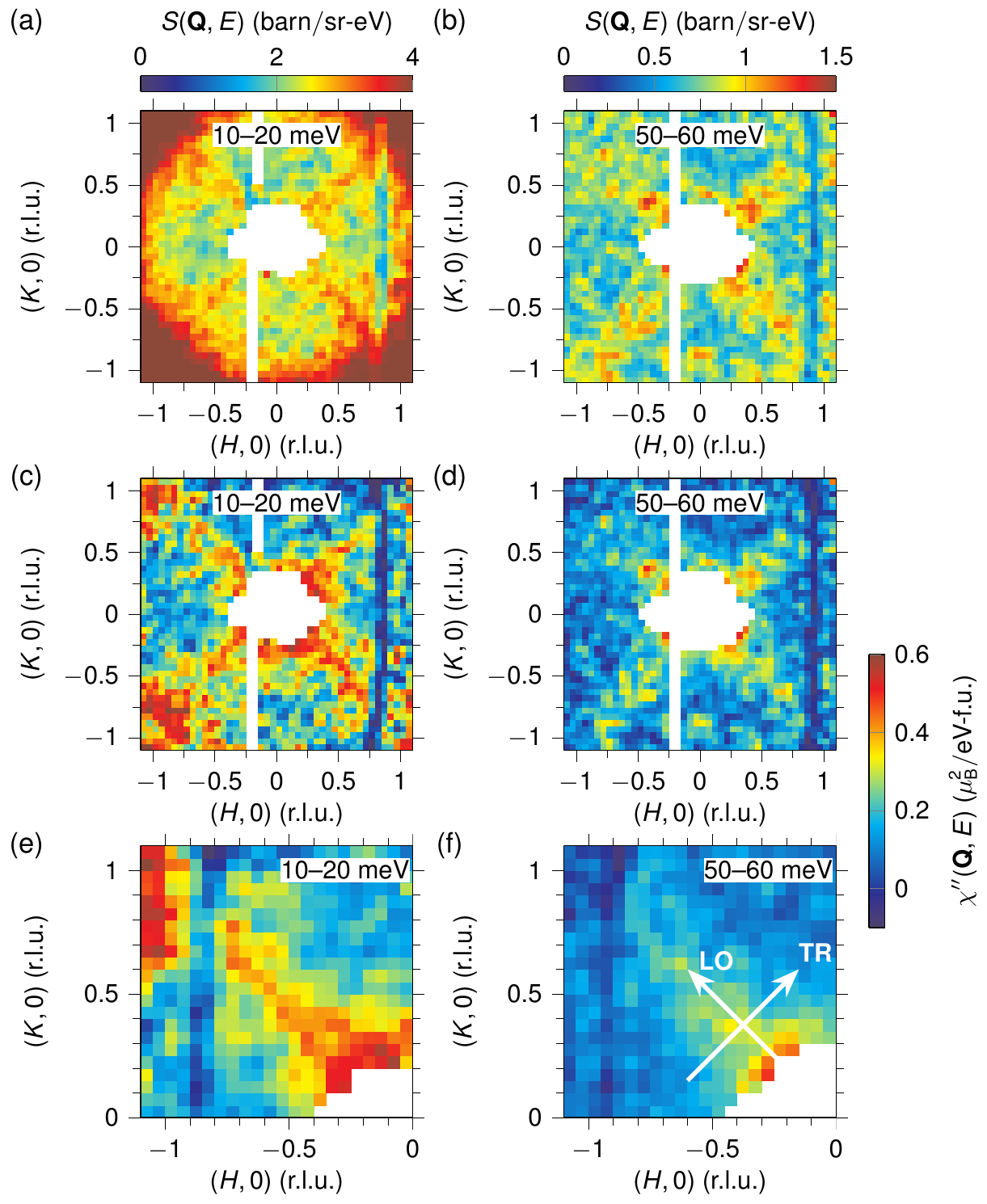}
	\caption{  \label{Fig:E_Slices_Co2} (a),(b) Constant-energy $E$ slices of inelastic-neutron-scattering cross section $S(\mathbf{Q},E)$ for the $(HK)$ reciprocal-lattice plane at $T=5.5$~K integrated over (a) $E=10$ to $20$~meV and (b) $50$ to $60$~meV. (c),(d)  Data corresponding to (a) and (b), respectively, plotted as the imaginary part of the magnetic susceptibility $\chi^{\prime\prime}$ after the isotropic background subtraction described in the main text.  The magnetic form factor for Co$^{2+}$ has been used.  (e),(f)  Data corresponding to (c) and (d), respectively, after averaging over symmetry-equivalent quadrants.  Data in (a), (c), and (e) are for an incident neutron energy of $E_{\text{i}}=75$~meV and data in (b),(d), and (f) are for  $E_{\text{i}}=125$~meV.  The transverse (TR) $[HH]$ and longitudinal (LO) $[-KK]$ directions are indicated in (f). }  
\end{figure}

For comparison to the $\chi^{\prime\prime}$ data  in the main text, for which we used a Co$^{1+}$ form factor, Fig.~\ref{Fig:E_Slices_Co2} shows constant-energy slices of INS data using $f(Q)$ for Co$^{2+}$ to determine $\chi^{\prime\prime}(\mathbf{Q},E)$.  Importantly, the quasi-$1$D spin fluctuations and very weak intensity (i.e.\ a small $\mu_{\text{fluct}}$) are unaffected.  However, using Co$^{2+}$ makes the longitudinally extended scattering decrease more steeply along the LO direction than for $\chi^{\prime\prime}$ calculated using Co$^{1+}$.  Fitting such data to the diffusive model gives   $\mu_{\text{fluc}}^2=0.017(1)~\mu_{\text{B}}^2/\text{f.u.}$, $\eta=-1.6(1)$, $\xi/a=0.71(5)$, $\chi^{\prime}(\mathbf{Q}_{\bm{\tau}},0)=0.00161(9)~\mu_{\text{B}}^2/\text{meV-f.u.}$, and $\Gamma=13(1)$~meV.  The lower value of $\eta$ than for the Co$^{1+}$ case is likely due to the vanishing of intensity below the level of our detection with increasing $Q$ along the LO direction.  Taking this into account, we do not believe that these data indicate  that Ca(Co$_{0.85}$Fe$_{0.15}$)$_{2}$As$_{2}$ is significantly less frustrated than CaCo$_{1.86}$As$_{2}$.

\section{Density Functional Theory Calculations}
Density functional theory (DFT) calculations were performed using a full-potential linear augmented plane wave (FP-LAPW) method, as implemented in \textsc{wien2k}~\cite{wien2k}.
We employed the generalized gradient approximation using the exchange-correlation functional of Perdew, Burke, and Ernzerhof~\cite{Perdew_1996}.  The muffin-tin (MT) radii $R_\text{MT}= 2.3$, $2.1$, and $2.1$~{a.u.}\ were used for Ca, Co, and As, respectively.  To generate the self-consistent potential and charge, we employed $R_\text{MT}K_\text{max}=8.0$, where $R_\text{MT}$ is the smallest muffin-tin radius, and $K_\text{max}$ is the plane-wave cutoff.  The calculations were iterated until the total energy differences were smaller than $0.01$~mRy.

The primitive cell contains one formula unit (f.u.), and experimental lattice parameters and $z_\text{As}$ (the $z$-coordinate of As atom) have been adopted \cite{Anand_2014,Jayasekara_2017}.  We chose $334$ $k$-points in the irreducible Brillouin zone (IBZ) for the self-consistent calculation and $726$ $k$-points for the densities of state (DOS) calculation.  [The energy dependent DOS is denoted $\rho(E)$ in the main text.]

To validate the rigid-band picture of Fe doping, we also investigated how the DOS near the Fermi energy $E_\text{F}$ evolves with Fe doping using supercell calculations.  We used supercells that correspond to a $2\times2\times1$ superstructure of the conventional CaCo$_2$As$_2$  ($2$~f.u.) unit cell.  The supercell consists of $40$ atoms ($16$ Co atoms); we substituted various numbers of Co atoms with Fe atoms to simulate different Fe doping concentrations. $84$ $k$-points were used in the IBZ for the supercell calculations. 

\begin{figure}[]
	\includegraphics[width=1.0\linewidth]{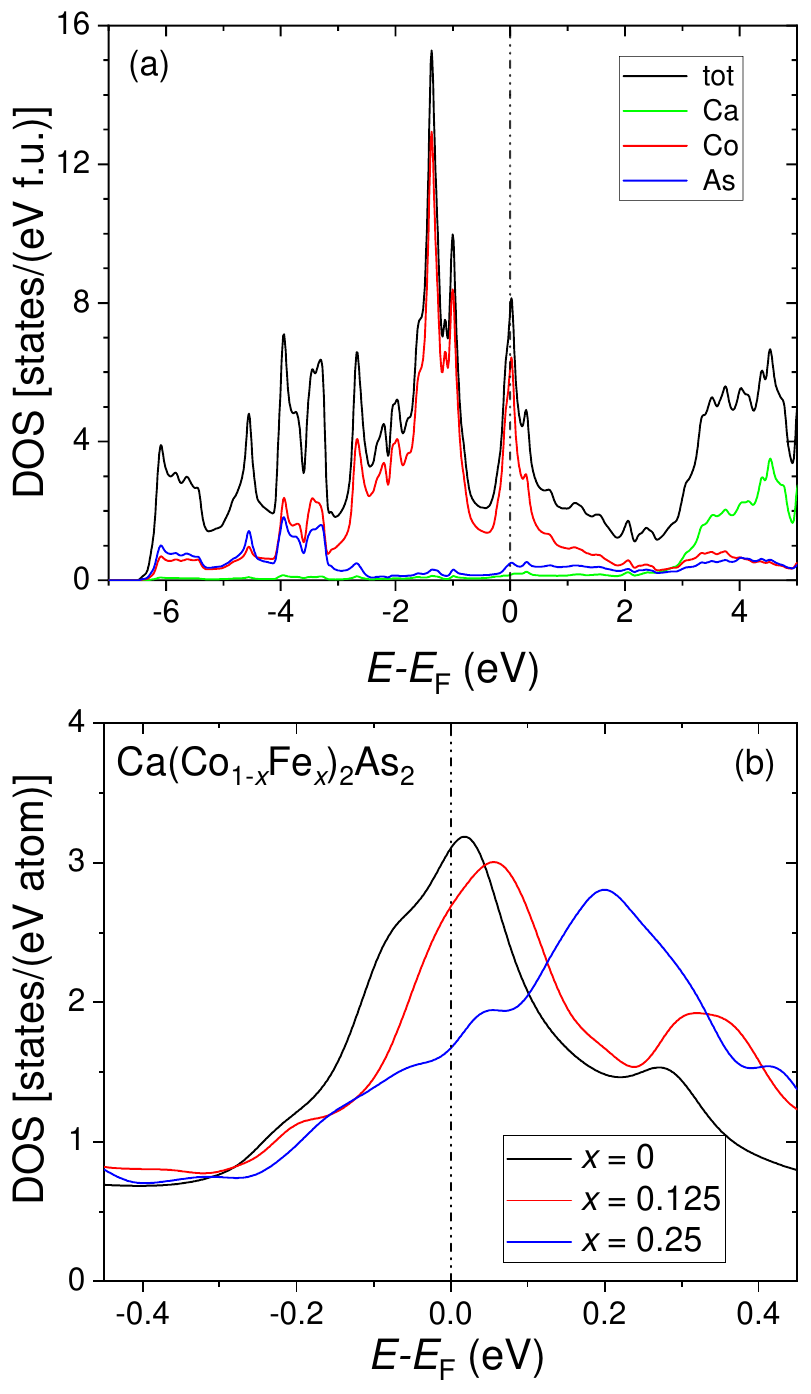}
	\caption{ \label{Fig:DFT} (a) The total (tot) and partial electronic density of states  ($\text{DOS}$) versus energy $E$ for CaCo$_2$As$_2$ from density functional theory calculations.  $E_{\text{F}}$ is the Fermi energy. (b) The partial DOS projected on $3d$ atoms in Ca(Co$_{1-x}$Fe$_{x}$)$_2$As$_2$ with $x=0$, $0.125$, and $0.25$}
\end{figure}

Figure~\ref{Fig:DFT}(a) shows the scalar-relativistic total ($\text{tot}$) DOS and partial DOS versus energy $E$ projected on individual atomic sites in nonmagnetic CaCo$_2$As$_2$.  A sharp peak dominated by carriers with Co orbital character crosses $E_{\text{F}}$.   This means that subtle changes to $E_{\text{F}}$ will tune the $\text{DOS}$ at $E_{\text{F}}$. Similar results from DFT calculations including  dynamical mean-field theory for the PM state of CaCo$_2$As$_2$ find a flat conduction band with $3d_{x^2-y^2}$ orbital character lying close to a Van Hove singularity at the corners of the Brillouin zone, as well as strong FM fluctuations \cite{Mao_2018}.  The flat conduction band appears around the $M$ reciprocal-lattice point.  These findings point to the occurrence of flat-band FM and long-wavelength FM fluctuations \cite{Tasaki_1998}. 

Figure~\ref{Fig:DFT}(b) shows the average partial DOS versus $E$ projected on $3d$ atoms in Ca(Co$_{1-x}$Fe$_{x}$)$_2$As$_2$ with $x=0$, $0.125$, and $0.25$.  These supercell calculation results show that hole doping shifts $E_\text{F}$ away from the DOS peak, validating the rigid-band picture.  As discussed in the main text, within Stoner theory such a decrease can tune the strength of the ferromagnetism and, in turn, the magnetic susceptibility \cite{Stoner_1938, Moriya_1985,Takahashi_2013}. 

\section{Low-Temperature Heat Capacity of $\bm{x=0.15}$}
\begin{figure}[]
	\includegraphics[width=1.0\linewidth]{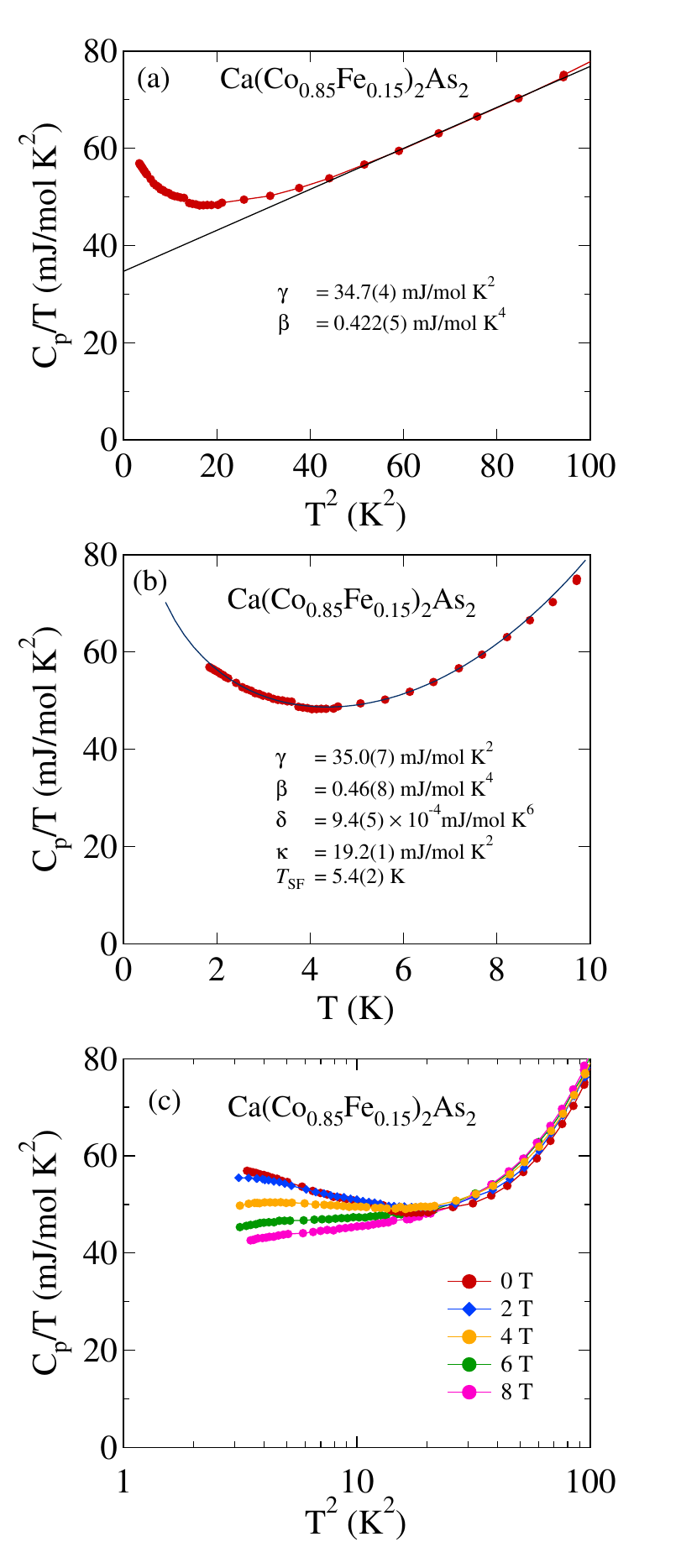}
	\caption{ \label{Fig:Cp_x0p15} Heat capacity at constant pressure divided by temperature $C_{\text{p}}/T$ for $x=0.15$ and zero applied magnetic field $H$ plotted versus $T^2$ (a) and $T$ (b). (c) $C_{\text{p}}/T$ versus $T^2$ for various values of $H$ plotted on a logarithmic horizontal axis.  Fits to the data in (a) and (b) are discussed in the text. }
\end{figure}

Heat capacity at constant pressure $C_{\text{p}}$ measurements were made on a $x=0.15$ single-crystal sample in a Quantum Design, Inc., Physical Properties Measurement System using a semi-adiabatic heat pulse technique.  Measurements were made down to a temperature of $T=2$~K and in applied magnetic fields up to $H=8$~T.

Figure~\ref{Fig:Cp_x0p15}(a) shows $C_{\text{p}}/T$ for $H=0$~T plotted versus $T^2$ and a fit to $C_{\text{p}}/T=\gamma+\beta T^2$, where $\gamma=34.7(4)~\text{mJ/mol-K}^2$ is the Sommerfeld coefficient and $\beta=0.422(5)~\text{mJ/mol-K}^4$ is the coefficient of the Debye term \cite{Kittel_2004}.  $\gamma$ is enhanced more than the value of $27(1)~\text{mJ/mol-K}^2$ found for $x=0$. $\beta$ is somewhat lower than the value of $1.008~\text{mJ/mol-K}^4$ determined for $x=0$ \cite{Anand_2014}, but this is likely due to the fitted range including contributions from quantum critical fluctuations.

The upturn in $C_{\text{p}}/T$ at low $T$ is due to non-Fermi liquid behavior attributed to critical fluctuations associated with a nearby quantum critical point.  To account for the upturn, the curve in Fig.~\ref{Fig:Cp_x0p15}(b) shows a fit to
\begin{equation}
	\frac{C_{\text{p}}}{T}=\gamma+\beta T^2+\delta T^4+\kappa\ln{\left(\frac{T}{T_{\text{SF}}}\right)}\ ,
\end{equation}
where $\delta$ accounts for higher-order contributions from the lattice and the final term accounts for ferromagnetic spin fluctuations with strength $\kappa$ and an energy scale related to a characteristic spin-fluctuation temperature $T_{\text{SF}}$ \cite{Sangeetha_2019}.  This final term has been associated with quantum fluctuations in, for example, Ni$_{x}$Pd$_{1-x}$ \cite{Nicklas_1999} and YFe$_{2}$Al$_{10}$ \cite{Wu_2014}.  The suppression of the low-$T$ upturn in $C_{\text{p}}/T$  with increasing $H$ shown in Fig.~\ref{Fig:Cp_x0p15}(c) is consistent with the upturn being due to ferromagnetic spin fluctuations.


\providecommand{\noopsort}[1]{}\providecommand{\singleletter}[1]{#1}%

\end{document}